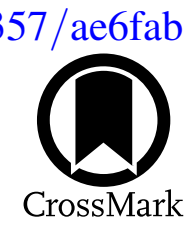

# The Variation of Circumstellar Parameters through Early Stellar Evolution

Michael Connelley[1,6], Christian Flores[2,3,4], and Bo Reipurth[1,5]
[1] University of Hawaii at Mānoa, 640 North Aohoku Place, Hilo, HI 96720, USA; mconnell@hawaii.edu
[2] Instituto de Estudios Astrofísicos, Facultad de Ingeniería y Ciencias, Universidad Diego Portales, Av. Ejército 441, Santiago, Chile
[3] Departamento de Física, Universidad de Santiago de Chile, Av. Victor Jara 3659, Santiago, Chile
[4] Millennium Nucleus on Young Exoplanets and their Moons (YEMS), Santiago, Chile
[5] Planetary Science Institute, 1700 East Fort Lowell Road, Suite 106, Tucson, AZ 85719, USA

## Abstract

We report on the evolution of extinction, mass accretion rate, and spectral index from analysis of SpeX observations and AllWISE spectral energy distributions, compared to stellar parameters (mass, age, and veiling) derived from analysis of previously reported and newly analyzed iSHELL observations of the same young stellar objects (YSOs). By using stellar gravity as an age indicator, we are able to infer age without any assumptions about the circumstellar environment. Our sample consists of 109 Class I, II, and III YSOs. The spectral index broadly decreases with time but with sufficient scatter that Class I–III YSOs can all be found at the same age. There is a wide range of extinction ($A_V$) from nearly 0 to >30 mag for ages up to 5 Myr, showing no clear trend with age. The mass accretion rate declines from a median of $10^{-8}\,M_\odot\,\mathrm{yr}^{-1}$ at ∼1 Myr to $<10^{-10}\,M_\odot\,\mathrm{yr}^{-1}$ at >10 Myr, with a scatter of the mass accretion rate of ±1 dex. Accretion seems to effectively end by ∼5 Myr, by which time the median mass gained is 0.05 $M_\odot$. We find a wide range of veiling ($r_K$) from zero to three (limited by our need to detect photospheric lines) for Class I and II YSOs, with the median veiling declining after ∼2 Myr. Class I YSOs appear to be physically very similar to Class II YSOs, with similar accretion rates and overlapping age and mass distributions. There are no objects with a high spectral index and low extinction, suggesting that once starved of circumstellar material, the spectral index rapidly declines.

## 1. Introduction

### 1.1. Age Indicators For Young Stars

Several commonly used age indicators for young stars trace the circumstellar environment and are unrelated to the central star itself. The class of T Tauri stars was originally defined in part by emission lines and proximity to dark clouds (A. H. Joy 1945). There is often the underlying assumption that higher IR excess, accretion, or extinction indicate a younger star. The Class I, II, and III sequence is understood to be an evolutionary sequence as defined by C. J. Lada (1987). Since this classification is purely empirical, extinction from the envelope or an edge-on disk can affect the spectral energy distribution (SED) slope. Also, IR excess, accretion, and extinction are each known to be variable (M. S. Connelley & T. P. Greene 2014). Ejections can make previously embedded stars optically visible (B. Reipurth 2000; B. Reipurth et al. 2010), and T Tauri N is itself likely to have been recently ejected from the embedded Class I T Tauri S binary, becoming optically visible in the process (C. Flores et al. 2020). T. P. Robitaille et al. (2006) presented the concept of stages of star formation, based on physical properties such as disk mass and mass accretion rate rather than the SED slope.

[6] Staff astronomer at the Infrared Telescope Facility, which is operated by the University of Hawaii under contract 80HQTR24DA010 with the National Aeronautics and Space Administration.

While accretion is a fundamental characteristic of young stars, it is difficult to use it as an age indicator due to its significant variability. Young stars experience eruptions, such as during EXor or FUor events (L. Hartmann & S. J. Kenyon 1996), where the accretion rate is much higher than normal. Hα equivalent width (EW) is used to classify an object as a classical or weak-line T Tauri star, and weak-line T Tauri stars (often associated with Class III objects) are generally assumed to be older. However, the Hα EW is variable, and objects near the threshold frequently cross between classical and weak-line T Tauri stars (C. Bertout & J. Bouvier 1989).

The color–magnitude diagram, with a comparison to theoretical evolutionary tracks, is commonly used to estimate the ages of young stars. However, this approach presents a number of problems. Different theoretical tracks disagree on the physical properties of a star (i.e., age and mass) for a given set of observations. Comparison with evolutionary tracks typically requires determining the $T_{\mathrm{eff}}$ of the star. M. A. Gully-Santiago et al. (2017) and C. Flores et al. (2022) found that the temperature inferred from spectroscopy depends on the wavelength chosen for the observation, and it is likely that neither optical nor near-IR observations yield an accurate $T_{\mathrm{eff}}$. Colors can be used to estimate $T_{\mathrm{eff}}$, but both colors and luminosity are affected by extinction, which is often difficult to accurately measure. Furthermore, accretion effects must also be accurately measured as accretion adds to the luminosity of the star and can also affect the color. These effects are worse for higher accreting and more deeply embedded young stellar objects (YSOs), where the uncertainties involved in estimating the extinction and accretion effects can be substantial.

There are several methods that are used to trace the ages of stars over 10 Myr. While gyrochronology works for older stars

(age > 100 Myr) in their spin-down phase (S. A. Barnes 2007), it does not work for Class I–II YSOs since the effects that make gyrochronology effective have not yet had the time to be the dominant factor in the spin rates of such young stars. F. Gallet & J. Bouvier (2013) show a spread in rotation rates of ∼1 dex until stars get to ∼100 Myr. X-rays and chromospheric activity have also been used as a sign of youth. T. Preibisch & E. D. Feigelson (2005) found a mild decay in X-ray luminosity versus age from 0.1 to 10 Myr in the Orion Nebula Cluster. However, this result was only for optically visible stars, and does not include heavily obscured objects. Furthermore, the scatter (FWHM) is about a factor of 10 in X-ray luminosity. Lithium abundance has also been used as a youth indicator. However, lithium's 6708 Å line is not detectable in deeply embedded young stars, and the line's presence or absence can only give an upper limit or lower limit on an individual star's age, respectively (D. R. Soderblom et al. 2014).

### *1.2. Gravity as an Age Indicator*

The surface gravity of an isolated star depends only on its mass and radius via the relation $G = MR^{-2}$ (in solar units, neglecting rotation effects). C. Hayashi (1961) showed that fully convective contracting stars evolve on the H-R diagram at nearly constant temperature, and this is supported by theoretical models of young stellar evolution (I. Baraffe et al. 2015; G. A. Feiden 2016). Since the isochrones are roughly parallel with lines of constant gravity, gravity is expected to be an indicator of stellar age. While different models disagree on the conversion from gravity to age, they do agree that gravity monotonically increases with time until the interior of the star starts to become radiative. Thus, while using gravity to establish absolute age is model dependent, gravity is an independent indicator of relative age for young stars on the Hayashi track. Gravity as an age indicator has the advantage that it is a physical property of the central star itself and is independent of distance (or distance uncertainties), circumstellar environment, line-of-sight extinction, or accretion effects.

The idea to use gravity, in conjunction with evolutionary models, to infer the ages of young stars dates back at least 30 yr, and in that time visible light and IR medium- and high-resolution spectroscopy has been used. R. P. Schiavon et al. (1995) used high-resolution visible light spectra to estimate $T_{\rm eff}$ and log($g$) for T Tauri stars, yielding log($g$) values quite close to C. Flores et al. (2022) and G. W. Doppmann & D. T. Jaffe (2003). They used the CO band heads, which are sensitive to luminosity class and hence gravity, but CO has the disadvantage of possibly being affected by the circumstellar environment. G. W. Doppmann et al. (2005) used high-resolution $K$-band spectra and found that the gravity of Class I and flat-spectrum objects is similar to Class II YSOs and have ages that span $10^7$ yr. A. R. Lyo et al. (2004) used visible light spectra of stars in the $\eta$ Cha group. Their line strength indices enabled them to show that the gravities of these stars lie between dwarfs and giants. C. L. Slesnick et al. (2006) also used visible light spectra and gravity sensitive lines (e.g., Na near 820 nm). They were able to distinguish between field stars, 5 Myr, and 1 Myr old stars with spectral types later than M2. Y. Takagi et al. (2011) used high-resolution $K$-band spectra near the 2.2 $\mu$m Na doublet to measure the gravities of four YSOs, using the relative depths of the Sc and Na lines as a tracer of gravity. Y. Yao et al. (2018) used high-resolution $H$-band spectra from the APOGEE survey to probe the disk frequency versus age for ∼2000 YSOs of the Orion A cloud, NGC 1333, and IC 348.

The gravity of YSOs must be precisely measured since YSOs are expected to have log($g$) values over the narrow range from 3.5 to 4.2. Doing this in the IR is preferable due to the extinction common among young stars. A new generation of high-resolution and high-bandwidth near-IR spectrographs, such as IGRINS and iSHELL, enable many photospheric lines to be observed at the same time. A series of papers from R. López-Valdivia et al. (2021) and C. Flores et al. (2019) has modeled the line profiles while simultaneously varying several stellar parameters (e.g., temperature, gravity, projected rotation velocity, radial velocity, metallicity, and magnetic field strength), taking advantage of the bandwidth of these new instruments.

### *1.3. Road Map*

The sample selection, observations, and biases are the same as for M. S. Connelley et al. (2026), and are described in Section 2. In Section 3 we describe the evolution of extinction, mass accretion, and veiling. In Section 4 we discuss the implications of our findings of the circumstellar evolution, and our conclusions are in Section 5.

## 2. Sample and Observations

### *2.1. Sample*

This paper continues the analysis of M. S. Connelley et al. (2026), and uses the same sample. Our sample of Class I, Class II, and Class III stars was selected to understand the evolution of the magnetic fields of young stars, and to compare spectroscopically derived masses against dynamical masses (C. Flores et al. 2022). To briefly summarize, we selected Class I YSOs brighter than $K = 11$ (to get a signal-to-noise ratio (S/N) ≈ 100 in a reasonable amount of time) and with low veiling (such that resolved photospheric absorption line shapes are measurable in a high-resolution spectrum, corresponding roughly to $r_K < 3$) primarily from M. S. Connelley & T. P. Greene (2010) and G. W. Doppmann et al. (2005; see C. Flores et al. 2024 for more details). We chose well studied classical T Tauri Stars from the nearby Taurus and Ophiuchus star-forming regions with dynamical masses from the Atacama Large Millimeter/submillimeter Array (ALMA; L. A. Cieza et al. 2019; M. Simon et al. 2019). The sample of T Tauri stars and their physical properties are presented in C. Flores et al. (2022). We selected Class III YSOs from the TW Hya association (∼10 Myr), the $\beta$ Pictoris moving group (S. Messina et al. 2017; ∼25 Myr), the AB Doradus group (B. Zuckerman et al. 2004), and the Pleiades (J. Bouvier et al. 2018; ∼100 Myr). We include the results presented by R. López-Valdivia et al. (2023) of T Tauri stars for the discussion of mass and SED evolution, as we do not have $JHK$ near-IR spectra of these targets necessary to probe the evolution of mass accretion tracers.

### *2.2. Known Biases*

The most important biases are the veiling of the Class I YSOs and disk masses of the Class II YSOs. As previously mentioned, our sample of Class I YSOs was limited to objects with sufficiently low veiling ($r_K < 3$) that the photospheric lines could be measured, and about one-half of Class I YSOs have veiling low enough to allow them to be used for this study. Low-veiling Class I YSOs have a lower spectral index overall than high-veiling YSOs, and thus may be more evolved as a whole.

Most of the Class II YSOs were also selected to have kinematic masses from ALMA observations, which rely on observing the kinematics of a spatially resolved circumstellar disk. Our sample of Class II YSOs is thus biased in favor of objects with larger and more massive disks. This likely accounts for the dearth of Class II YSOs in our sample with spectral index values between −1 and −2. There may be a correlation between disk mass and stellar mass (S. M. Andrews et al. 2013; C. F. Manara et al. 2023), so we may have a bias toward higher-mass Class II YSOs. Our limiting magnitude is $K \sim 11$, so we are limited to stars with masses $> 0.3\,M_{\odot}$. We selected stars cooler than ∼5000 K to be able to measure magnetic fields. As such, our sample contains no Herbig Ae/Be stars. These selection criteria constrain our Class II sample to a mass range of ∼0.3 to ∼1.5 $M_{\odot}$.

Brighter stars are easier to see regardless of age, so all groups would be biased toward brighter and hotter stars, which are presumably higher mass. Class II–III YSOs may be less affected by this bias since their low extinction allows us to observe fainter (and presumably lower-mass) stars. For stars descending the Hayashi track, as many of our Class I–II YSOs are, younger stars are more luminous because they are larger. The higher extinction that is characteristic of Class I YSOs may preferentially bias that sample toward brighter (and presumably higher-mass) stars, as many Class I YSOs are too faint for high-resolution spectroscopy. Our sample may also be biased toward lower-extinction Class I YSOs. Since we can only determine the physical properties of Class I YSOs with low veiling, and since low-veiling Class I YSOs tend to have lower spectral indices, our sample of Class I YSOs is biased toward lower spectral indices. The Class I YSOs bias for hotter objects could mask an increasing trend of mass versus age should Class I YSOs be less massive than Class II–III YSOs. The Class I YSO sample may be biased toward stars with higher accretion rates (such stars being generally brighter), however the selection of only Class I YSOs with low veiling likely counters this to some degree.

### *2.3. Observations*

Observations were carried out with the 3.2 m NASA Infrared Telescope Facility (IRTF) on Maunakea, Hawaii with SpeX and iSHELL. SpeX was used in the short cross-dispersed mode, covering 0.7–2.5 $\mu$m (J. T. Rayner et al. 2003). We used the 0$''\!.$5 wide slit, yielding a resolving power of $R = 1200$. Stars were nodded along the slit, with two exposures taken at each nod position in the usual ABBA beam switch pattern. An A0 telluric standard star was observed after at least every other science target for telluric correction, usually within 0.1 air masses of the target and within 1 hr. Our observing log is shown in Table 1. A thorium-argon lamp was observed for wavelength calibration and a quartz lamp for flat-fielding. An arc–flat calibration set was observed for each target–standard pair.

iSHELL (J. Rayner et al. 2016) was used in the K2 mode, covering from 2.08 to 2.38 $\mu$m, with the 0$''\!.$75 slit, yielding a resolving power of $R = 55{,}000$. The star is kept in the center of the slit, without nodding. The total exposure time was determined to get an $S/N \approx 100$ spectrum for each target, with individual exposures of 300–600 s. An A0 telluric standard star was observed after each science target. An arc–flat calibration set was observed for each target and another set was taken for each standard star, making sure each calibration set was done before moving the telescope.

The SpeX and iSHELL data were flat-fielded, extracted, and wavelength calibrated using `Spextool` (M. C. Cushing et al. 2004). Spectral line flux, EW, and FWHM were measured using the SPLOT routine in IRAF. The spectra were pseudoflux calibrated during telluric correction, using the $B$ and $V$ magnitudes of the A0 standard star and assuming the same throughput versus wavelength for both the telluric standard and the science target. However, this pseudoflux calibration can be affected by nonphotometric weather or changes in seeing. Final flux calibration was done using the Two Micron All Sky Survey (2MASS) $K$-band magnitude of each star.

The physical properties (temperature, log($g$), and veiling) of the Class II YSOs are from C. Flores et al. (2022), the physical properties of the Class I and flat-spectrum YSOs are from C. Flores et al. (2024), and the physical properties of the Class III YSOs are from C. Flores (2026, private communication). The process of determining stellar parameters and their uncertainties from high-resolution spectra is detailed in C. Flores et al. (2019). For the low-veiling Class I YSOs, we ignored the region around the 2.110 and 2.117 $\mu$m Al lines and 2.228 $\mu$m Ti line as the continuum in these regions is strongly affected by water. C. Flores et al. (2024) show that this change in the procedure has a negligible impact on the derived physical parameters. The derived properties of our sample are presented in Table 2, where we note the source of the physical parameters for each star in the last column of Table 1.

iSHELL observations from 2017 April were taken by T. Sullivan et al. (2019). C. Flores et al. (2024) retrieved the data from the IRTF archive, then reduced and modeled the data as described above. Other iSHELL observations were taken by C. Flores et al. (2020, 2022, 2024). All SpeX observations taken in 2017 and later are newly reported here. SpeX observations from 2006 to 2007 are from M. S. Connelley & T. P. Greene (2010).

## 3. Analysis

### *3.1. Stellar Parameters*

The process by which we determined the stellar parameters via high-resolution $K$-band spectroscopy is detailed in C. Flores et al. (2019). In summary, the reduced spectrum is first continuum normalized. MOOGStokes is used to generate a model spectrum (C. P. Deen 2013), by performing radiative transfer modeling through a magnetized stellar atmosphere. One input into MOOGStokes is the model stellar atmosphere, for which MARCS (B. Gustafsson et al. 2008) 1D LTE hydrostatic models are used. The temperature, gravity (log $g$), veiling (metallicity), rotation, microturbulence, and magnetic field strength are simultaneously varied via Markov Chain Monte Carlo to efficiently explore the parameter space. The papers that provide these values are referenced in Table 1. Hence temperature, log($g$), veiling, and rotation are direct outputs of the spectral modeling process.

Masses and ages are determined by comparing the temperature and gravity values to the G. A. Feiden (2026, private communication) evolutionary models. These models are similar to the G. A. Feiden (2016) magnetic evolutionary models we used previously, but now extend to 0.1 Myr age. Mass primarily depends on temperature, whereas age primarily depends on

**Table 1**
Sample Targets

| Object | Alternate Name | $\alpha$ (HH:MM:SS.S) | $\delta$ (DD:MM:SS) | $K$ (mag) | SpeX UT Date (YYMMDD) | iSHELL UT Date (YYMMDD) | Source |
|---|---|---|---|---|---|---|---|
| HIP 11437 A | SE | 02:27:29.2 | +30:58:25 | 7.1 | ⋯ | 180807 | 4 |
| HIP 11437 B | NW | 02:27:29.2 | +30:58:25 | 7.1 | ⋯ | 180807 | 4 |
| HIP 12545 | ⋯ | 02:41:25.9 | +05:59:18 | 7.1 | ⋯ | 180807 | 4 |
| HIP 12635 | ⋯ | 02:42:20.9 | +38:37:21 | 7.8 | ⋯ | 180807 | 4 |
| IRAS 03301+3111 | ⋯ | 03:33:12.8 | +31:21:24 | 9.0 | 181201 | 181206 | 3 |
| HIP 16563 A | NW | 03:33:13.5 | +46:15:27 | 6.4 | ⋯ | 181005 | 4 |
| ⋯ | ⋯ | ⋯ | ⋯ | ⋯ | ⋯ | **180831** | ⋯ |
| HIP 16563 B | SE | 03:33:13.5 | +46:15:27 | ⋯ | ⋯ | 181005 | 4 |
| ⋯ | ⋯ | ⋯ | ⋯ | ⋯ | ⋯ | **180831** | ⋯ |
| HIP 17695 | ⋯ | 03:47:23.3 | −01:58:20 | 6.9 | ⋯ | 181005 | 4 |
| EPIC 211068400 | ⋯ | 03:53:23.7 | +24:03:54 | 9.4 | ⋯ | 181018 | 4 |
| HIP 18859 | ⋯ | 04:02:36.7 | −00:16:08 | 4.2 | ⋯ | 181005 | 4 |
| IRAS 04108+2803 E | ⋯ | 04:13:54.7 | +28:11:31 | 11.1 | 070302 | 200216 | 3 |
| FM Tau | ⋯ | 04:14:13.6 | +28:12:49 | 8.8 | 171010 | 200109 | 2 |
| IRAS 04113+2758 S | ⋯ | 04:14:26.4 | +28:06:00 | 7.8 | 061101 | 200125 | 3 |
| FP Tau | ⋯ | 04:14:47.3 | +26:46:26 | 8.9 | 171010 | 171013 | 2 |
| CX Tau | ⋯ | 04:14:47.9 | +26:48:11 | 8.8 | 171010 | 171013 | 2 |
| CY Tau | ⋯ | 04:17:33.7 | +28:20:47 | 8.6 | 171010 | 171013 | 2 |
| BP Tau | ⋯ | 04:19:15.8 | +29:06:27 | 7.7 | 170120 | 171106 | 2 |
| IRAS 04181+2665 M | ⋯ | 04:21:10.4 | +27:01:37 | 11.1 | 181201 | 190113 | 3 |
| IRAS 04181+2665 S | ⋯ | 04:21:11.5 | +27:01:09 | 10.3 | 181201 | 190101 | 3 |
| DE Tau | ⋯ | 04:21:55.6 | +27:55:06 | 7.8 | 200127 | 191116 | 2 |
| T Tauri N | ⋯ | 04:21:59.4 | +19:32:06 | 5.3 | 191009 | 191006 | 1 |
| IP Tau | ⋯ | 04:24:57.1 | +27:11:57 | 8.4 | 171124 | 180114 | 2 |
| FV Tau A | ⋯ | 04:26:53.5 | +26:06:54 | 8.9 | 171115 | 180101 | 2 |
| DG Tau | IRAS 04240+2559 | 04:27:04.7 | +26:06:16 | 7.0 | 070303 | 200125 | 3 |
| GV Tau S | Haro 6-10S | 04:29:23.7 | +24:33:00 | 8.1 | 201018 | 191121 | 3 |
| IQ Tau | ⋯ | 04:29:51.6 | +26:06:45 | 7.8 | 171115 | 200119 | 2 |
| DK Tau A | ⋯ | 04:30:44.3 | +26:01:24 | 7.1 | 171124 | 180113 | 2 |
| DK Tau B | ⋯ | 04:30:44.4 | +26:01:24 | ⋯ | 171124 | 180113 | 2 |
| HK Tau A | ⋯ | 04:31:50.6 | +24:24:18 | 8.6 | 191227 | 180114 | 2 |
| V710 Tau N | ⋯ | 04:31:57.8 | +18:21:37 | 8.7 | 191227 | 180114 | 2 |
| V710 Tau S | ⋯ | 04:31:57.8 | +18:21:37 | ⋯ | 191227 | 180114 | 2 |
| Haro 6-13 | IRAS 04292+2422 E | 04:32:15.4 | +24:29:00 | 8.1 | 061101 | 191119 | 3 |
| IRAS 04295+2251 | LDN 1536 IRS | 04:32:32.1 | +22:57:27 | 10.1 | 181201 | 190101 | 3 |
| GK Tau A | ⋯ | 04:33:34.6 | +24:21:06 | 7.5 | 171124 | 191017 | 2 |
| DL Tau | ⋯ | 04:33:39.1 | +25:20:38 | 8.0 | 180228 | 200125 | 2 |
| DM Tau | ⋯ | 04:33:48.7 | +18:10:10 | 9.5 | 171115 | 200125 | 2 |
| CI Tau | ⋯ | 04:33:52.0 | +22:50:30 | 7.8 | 171115 | 180113 | 2 |
| AA Tau | ⋯ | 04:34:55.4 | +24:28:53 | 8.1 | 200127 | 180131 | 2 |
| HO Tau | ⋯ | 04:35:20.2 | +22:32:14 | 9.7 | 171115 | 180115 | 2 |
| DN Tau | ⋯ | 04:35:27.4 | +24:14:59 | 8.0 | 171115 | 180113 | 2 |
| HP Tau | ⋯ | 04:35:52.8 | +22:54:23 | 7.6 | 190930 | 201023 | 2 |
| Haro 6-28 | V1026 Tau | 04:35:56.9 | +22:54:36 | 9.5 | 190930 | 161016 | 3 |
| LkCa 15 | ⋯ | 04:39:17.8 | +22:21:03 | 8.2 | 180227 | 180112 | 2 |
| Haro 6-33 | ⋯ | 04:41:38.8 | +25:56:26 | 9.2 | 201018 | 201023 | 2 |
| GO Tau | ⋯ | 04:43:03.1 | +25:20:19 | 9.3 | 191227 | 180114 | 2 |
| DS Tau | ⋯ | 04:47:48.6 | +29:25:11 | 8.0 | 191227 | 191017 | 2 |
| UY Aur NE | ⋯ | 04:51:47.4 | +30:47:14 | 7.3 | ⋯ | 201023 | 2 |
| IRAS 04489+3042 | ⋯ | 04:52:06.7 | +30:47:18 | 10.4 | ⋯ | 200215 | 3 |
| GM Aur | ⋯ | 04:55:11.0 | +30:21:59 | 8.3 | ⋯ | 171106 | 2 |
| V347 Aur | IRAS 04530+5126 | 04:56:57.0 | +51:30:51 | 7.8 | 070302 | 180831 | 3 |
| IRAS 04591−0856 | ⋯ | 05:01:29.6 | −08:52:17 | 10.6 | 181201 | 181231 | 3 |
| IRAS 05379−0758(2) | ⋯ | 05:40:20.31 | −07:56:25 | 10.9 | 181201 | 200308 | 3 |
| IRAS 05555-1405(4) | ⋯ | 05:57:49.5 | −14:05:34 | 10.2 | 181201 | 190101 | 3 |
| TWA 6 | ⋯ | 10:18:28.7 | −31:50:03 | 8.0 | 200131 | 181231 | 4 |
| TWA 7 | ⋯ | 10:42:30.1 | −33:40:16 | 6.9 | 200131 | 191218 | 4 |
| TWA 3 B | ⋯ | 11:10:27.8 | −37:31:53 | ⋯ | 200131 | 190101 | 4 |
| TWA 13 N | ⋯ | 11:21:17.2 | −34:46:45 | 7.5 | 200131 | 190203 | 4 |
| TWA 13 S | ⋯ | 11:21:17.4 | −34:46:50 | 7.5 | 200131 | 190203 | 4 |
| TWA 4 | ⋯ | 11:22:05.3 | −24:46:40 | 5.6 | ⋯ | 210129 | 4 |
| TWA 8 A | ⋯ | 11:32:41.3 | −26:51:56 | 7.4 | 191227 | 190101 | 4 |
| TWA 9 A | ⋯ | 11:48:24.2 | −37:28:49 | 7.8 | 200131 | 190203 | 4 |

**Table 1**
(Continued)

| Object | Alternate Name | $\alpha$ (HH:MM:SS.S) | $\delta$ (DD:MM:SS) | $K$ (mag) | SpeX UT Date (YYMMDD) | iSHELL UT Date (YYMMDD) | Source |
|---|---|---|---|---|---|---|---|
| TWA 9 B | ⋯ | 11:48:23.7 | −37:28:49 | 9.2 | 200131 | 190203 | 4 |
| TYC 915-1391-1 | ⋯ | 14 25 55.9 | +14:12:10 | 7.3 | 200131 | 190203 | 4 |
| ⋯ | ⋯ | ⋯ | ⋯ | ⋯ | ⋯ | **210129** | ⋯ |
| GSS 26 | ODISEA 71 | 16:26:10.3 | −24:20:55 | 8.5 | ⋯ | 190709 | 2 |
| DoAr 25 | IRAS 16234−2436 | 16:26:23.7 | −24:43:14 | 7.9 | ⋯ | 180723 | 2 |
| ⋯ | ODISEA 89 | ⋯ | ⋯ | ⋯ | ⋯ | ⋯ | ⋯ |
| [GY92] 33 | ODISEA 94 | 16:26:27.5 | −24:41:54 | 10.0 | 210625 | 170428 | 3 |
| ODISEA 107 W | GSS 37 | 16:26:42.8 | −24:20:30 | 7.9 | ⋯ | 190709 | 2 |
| Elias 2-27 | GSS 39 | 16:26:45.0 | −24:23:08 | 9.0 | ⋯ | 200706 | 2 |
| SR 24 S | Haro 1-7 | 16:26:58.5 | −24:45:37 | 7.1 | 210625 | 170426 | 3 |
| [GY92] 235 | ⋯ | 16:27:13.8 | −24:43:32 | 10.0 | ⋯ | 170427 | 3 |
| WL 20 E | ⋯ | 16:27:15.7 | −24:38:43 | 9.5 | ⋯ | 170428 | 3 |
| WL 20 W | ⋯ | 16:27:15.7 | −24:38:43 | 9.5 | ⋯ | 170427-8 | 3 |
| WLY 2-42 | YLW 13B | 16:27:21.5 | −24:41:43 | 8.5 | ⋯ | 170427 | 3 |
| Elias 2-32 | ODISEA 167 | 16:27:28.4 | −24:27:21 | 10.1 | 200819 | 200821 | 3 |
| Elias 2-33 | ODISEA 169 | 16:27:30.2 | −24:27:44 | 9.0 | 200819 | 200706 | 3 |
| ⋯ | VSSG17 | ⋯ | ⋯ | ⋯ | ⋯ | ⋯ | ⋯ |
| ROX 25 | YLW 18 | 16:27:33.1 | −24:41:15 | 7.8 | ⋯ | 200706 | 2 |
| DoAr 33 | ODISEA 185 | 16:27:39.0 | −23:58:18 | 8.2 | ⋯ | 190709 | 2 |
| ROX 27 | ODISEA 186 | 16:27:39.4 | −24:39:15 | 8.5 | ⋯ | 190709 | 2 |
| YLW 58 | ⋯ | 16:28:16.5 | −24:36:58 | 9.3 | ⋯ | 200706 | 2 |
| DoAr 43 | ODISEA 250 | 16:31:30.9 | −24:24:40 | 7.9 | ⋯ | 190709 | 2 |
| Haro 1-16 | ODISEA 252 | 16:31:33.4 | −24:27:37 | 7.6 | 180801 | 200825 | 2 |
| IRAS 16285-2358 | ODISEA 253 | 16:31:33.8 | −24:04:47 | 10.0 | 200819 | 200821 | 3 |
| IRAS 16288−2450 W2 | ODISEA 263 SW | 16:31:52.1 | −24:56:16 | 8.7 | 190516 | 180723 | 3 |
| ⋯ | ISO-Oph 204 SW | ⋯ | ⋯ | ⋯ | ⋯ | ⋯ | ⋯ |
| HIP 81084 | ⋯ | 16:33:41.6 | −09:33:12 | 7.5 | 180801 | 180723 | 4 |
| WSB 82 | IRAS 16367−2356 | 16:39:45.4 | −24:02:04 | 7.6 | ⋯ | 180724 | 2 |
| ⋯ | ODISEA 283 | ⋯ | ⋯ | ⋯ | ⋯ | ⋯ | ⋯ |
| 2MASS J16430128−1754274 | ⋯ | 16:43:01.3 | −17:54:27 | 8.5 | ⋯ | 180723 | 4 |
| HIP 82688 | ⋯ | 16:54:08.1 | −04:20:25 | 6.4 | ⋯ | 180723 | 4 |
| [EC92] 92 | ⋯ | 18:29:57.7 | +01:12:52 | 10.5 | 190516 | 200706 | 3 |
| [EC92] 95 | ⋯ | 18:29:57.9 | +01:12:46 | 9.7 | 190516 | 190707 | 3 |
| [TS84] IRS 5 NE | ⋯ | 19:01:48.1 | −36:57:22 | 10.1 | ⋯ | 190707 | 3 |
| 2MASS J19102820−2319486 | ⋯ | 19:10:28.2 | −23:19:49 | 8.2 | 180624 | 180807 | 4 |
| TYC 6878-0195-1 | ⋯ | 19:11:44.7 | −26:04:09 | 7.4 | ⋯ | 180628 | 4 |
| IRAS 19247+2238(1) | ⋯ | 19:26:51.3 | +22:45:13 | 8.4 | 190624 | 191116 | 3 |
| IRAS 19247+2238(2) | ⋯ | 19:26:51.3 | +22:45:13 | 9.1 | 190703 | 190706 | 3 |
| 2MASSJ20013718−3313139 | ⋯ | 20:01:37.2 | −33:13:13 | 8.2 | ⋯ | 191015 | 4 |
| ⋯ | ⋯ | ⋯ | ⋯ | ⋯ | ⋯ | **200427** | ⋯ |
| 2MASSJ20055640−3216591 | ⋯ | 20:05:56.4 | −32:16:59 | 7.9 | ⋯ | 191015 | 4 |
| TYC 6349-0200-1 NW | AZ Cap | 20:56:02.7 | −17:10:54 | 7.0 | 180624 | 180807 | 4 |
| TYC 6349-0200-1 SE | AZ Cap | 20:56:02.7 | −17:10:54 | 7.0 | 180624 | 180807 | 4 |
| 2MASS J21100535−1919573 | ⋯ | 21:10:05.4 | −19:19:57 | 7.2 | 180624 | 180628 | 4 |
| HIP 110526 E | ⋯ | 22:23:29.1 | +32:27:34 | 6.1 | 180624 | 180724 | 4 |
| HIP 110526 W | ⋯ | 22:23:29.1 | +32:27:34 | ⋯ | 180624 | 180724 | 4 |
| HIP 113579 | ⋯ | 23:00:19.4 | −26:09:15 | 5.9 | 180624 | 180628 | 4 |

**Notes.** Sources of stellar properties from high-resolution IR spectroscopy:
1: C. Flores et al. (2020).
2: C. Flores et al. (2022).
3: C. Flores et al. (2024).
4: Flores, private communication.
Bold dates designate the data that we present and use to extract stellar parameters.

gravity. The uncertainties in the inferred masses were derived by using the upper and lower temperature limits into our interpolation code. Some stars are too hot or old to be on the Hayashi tracks (where stars evolve downward with increasing gravity with age, and have turned onto the Henyey track, where stars evolve with more constant gravity). We cannot reliably measure the ages of such stars, and they are noted as "Henyey" in Table 2. A more detailed discussion of how we convert from $\log(g)$ to ages is presented in M. Connelley et al. (2026). Stellar radius, later used to calculate the mass accretion rate, is derived using the measured surface gravity and inferred mass via $R = \sqrt{GM/g}$.

**Table 2**
Stellar and Circumstellar Properties

| Object | Age (Myr) | Mass ($M_\odot$) | Index | $A_V$ (mag) | Br$\gamma \times 10^{-15}$ (erg s$^{-1}$ cm$^{-2}$) | $\dot{M} \times 10^{-9}$ ($M_\odot$ yr$^{-1}$) | Veiling ($rK$) |
|---|---|---|---|---|---|---|---|
| 2M16430128 | $38.8^{+1.9}_{-1.9}$ | $0.66^{+0.01}_{-0.01}$ | $-2.65 \pm 0.10$ | no data | ⋯ | ⋯ | $0.05^{+0.01}_{-0.01}$ |
| 2M19102820 | $32.5^{+1.2}_{-1.2}$ | $0.42^{+0.01}_{-0.01}$ | $-2.60 \pm 0.10$ | $0.6^{+0.3}_{-0.6}$ | undetected | N/A | $0.00^{+0.01}_{-0.00}$ |
| 2M20013718 | $34.4^{+1.6}_{-0.0}$ | $0.63^{+0.01}_{-0.01}$ | $-2.67 \pm 0.09$ | no data | ⋯ | ⋯ | $0.04^{+0.03}_{-0.01}$ |
| 2M20055640 | $32.7^{+1.5}_{-1.4}$ | $0.62^{+0.02}_{-0.03}$ | $-2.65 \pm 0.09$ | no data | ⋯ | ⋯ | $0.12^{+0.06}_{-0.04}$ |
| 2M21100535 | $33.7^{+1.5}_{-0.0}$ | $0.61^{+0.01}_{-0.01}$ | $-2.62 \pm 0.10$ | $0.4^{+0.4}_{-0.2}$ | $1.5^{+0.6}_{-0.5}$ | $0.02^{+0.01}_{-0.01}$ | $0.06^{+0.00}_{-0.00}$ |
| AA Tau | $2.9^{+0.8}_{-0.7}$ | $0.55^{+0.05}_{-0.05}$ | $-0.55 \pm 0.27$ | $3.5^{+1.8}_{-0.8}$ | $43.5^{+20.9}_{-13.1}$ | $3.5^{+2.9}_{-1.5}$ | $0.99^{+0.05}_{-0.05}$ |
| BP Tau | $12.7^{+4.3}_{-3.0}$ | $0.74^{+0.03}_{-0.03}$ | $-0.78 \pm 0.28$ | $0.1^{+0.8}_{-0.1}$ | $112.8^{+20.0}_{-11.4}$ | $6.6^{+2.1}_{-1.3}$ | $1.36^{+0.07}_{-0.06}$ |
| CI Tau | $1.8^{+0.7}_{-0.7}$ | $0.83^{+0.06}_{-0.09}$ | $-0.70 \pm 0.21$ | $0.5^{+0.6}_{-0.4}$ | $639.0^{+53.7}_{-38.0}$ | $96^{+26}_{-20}$ | $1.87^{+0.08}_{-0.08}$ |
| CX Tau | $2.5^{+0.4}_{-0.5}$ | $0.47^{+0.03}_{-0.04}$ | $-0.55 \pm 0.35$ | $0.5^{+1.1}_{-0.4}$ | $25.6^{+5.4}_{-3.0}$ | $2.1^{+0.8}_{-0.4}$ | $0.26^{+0.03}_{-0.02}$ |
| CY Tau | $3.3^{+0.4}_{-0.4}$ | $0.41^{+0.04}_{-0.02}$ | $-1.19 \pm 0.19$ | $0.5^{+0.8}_{-0.5}$ | $24.3^{+4.4}_{-3.2}$ | $2.0^{+0.6}_{-0.4}$ | $0.58^{+0.04}_{-0.04}$ |
| DE Tau | $2.1^{+0.4}_{-0.4}$ | $0.43^{+0.03}_{-0.03}$ | $-0.80 \pm 0.29$ | $0.2^{+0.7}_{-0.2}$ | $54.6^{+8.7}_{-5.9}$ | $6.2^{+1.9}_{-1.3}$ | $1.15^{+0.04}_{-0.04}$ |
| DG Tau | $0.45^{+0.17}_{-0.08}$ | $0.83^{+0.05}_{-0.06}$ | $0.17 \pm 0.24$ | $0.8^{+1.0}_{-0.3}$ | $550^{+80}_{-40}$ | $125^{+53}_{-34}$ | $3.60^{+0.30}_{-0.38}$ |
| DK Tau A | $2.2^{+0.5}_{-0.5}$ | $0.46^{+0.03}_{-0.02}$ | $-0.71 \pm 0.18$ | $0.6^{+0.5}_{-0.4}$ | $91.0^{+15.1}_{-13.1}$ | $11^{+3.5}_{-2.8}$ | $1.90^{+0.08}_{-0.08}$ |
| DK Tau B | $2.8^{+0.5}_{-0.4}$ | $0.41^{+0.05}_{-0.01}$ | $-0.71 \pm 0.18$ | $1.8^{+1.0}_{-0.7}$ | $20.6^{+4.1}_{-3.0}$ | $1.7^{+0.6}_{-0.4}$ | $1.10^{+0.06}_{-0.04}$ |
| DL Tau | $2.3^{+0.6}_{-0.5}$ | $0.89^{+0.04}_{-0.04}$ | $-0.62 \pm 0.19$ | $0.0^{+0.9}_{-0.0}$ | $330^{+45.6}_{-13.4}$ | $37^{+10}_{-4.8}$ | $2.97^{+0.07}_{-0.08}$ |
| DM Tau | $4.6^{+0.3}_{-0.4}$ | $0.32^{+0.03}_{-0.01}$ | $-0.41 \pm 0.61$ | $0.6^{+1.4}_{-0.6}$ | $36.8^{+9.0}_{-5.0}$ | $3.4^{+1.2}_{-0.7}$ | $0.03^{+0.02}_{-0.01}$ |
| DN Tau | $1.9^{+0.4}_{-0.3}$ | $0.47^{+0.03}_{-0.02}$ | $-0.75 \pm 0.32$ | $0.1^{+0.6}_{-0.1}$ | $62.0^{+10.6}_{-6.9}$ | $7.1^{+2.2}_{-1.4}$ | $0.62^{+0.02}_{-0.02}$ |
| DoAr 25 | $1.1^{+0.4}_{-0.2}$ | $0.59^{+0.05}_{-0.05}$ | $-0.96 \pm 0.35$ | no data | ⋯ | ⋯ | $0.50^{+0.03}_{-0.03}$ |
| DoAr 33 | $3.1^{+1.2}_{-0.8}$ | $0.64^{+0.02}_{-0.03}$ | $-1.12 \pm 0.22$ | no data | ⋯ | ⋯ | $0.70^{+0.01}_{-0.02}$ |
| DoAr 43 | $1.1^{+0.4}_{-0.3}$ | $1.13^{+0.07}_{-0.06}$ | $-0.51 \pm 0.21$ | no data | ⋯ | ⋯ | $1.96^{+0.14}_{-0.17}$ |
| DS Tau | $2.1^{+0.9}_{-0.5}$ | $0.56^{+0.06}_{-0.05}$ | $-1.02 \pm 0.19$ | $0.5^{+0.4}_{-0.3}$ | $99.1^{+9.5}_{-9.0}$ | $11^{+2.8}_{-2.4}$ | $1.61^{+0.05}_{-0.05}$ |
| EC 92 | $0.10^{+0.06}_{-?}$ | $0.49^{+0.05}_{-0.04}$ | $0.92 \pm 0.29$ | $26.6^{+5.4}_{-6.6}$ | undetected | N/A | $0.91^{+0.1}_{-0.1}$ |
| EC 95 | $2.2^{+1.2}_{-1.1}$ | $1.21^{+0.12}_{-0.08}$ | $0.92 \pm 0.29$ | $35^{+2}_{-8}$ | undetected | N/A | $0.95^{+0.11}_{-0.37}$ |
| Elias 2-32 | $< 0.1$ | off grid | $0.14 \pm 0.25$ | $27^{+5}_{-5}$ | $20.4^{+24.6}_{-12.2}$ | $5.3^{+11}_{-3.8}$ | $0.48^{+0.06}_{-0.05}$ |
| Elias 2-33 | $0.1^{+0.5}_{-?}$ | $0.26^{+0.06}_{-0.06}$ | $0.06 \pm 0.41$ | $28^{+6}_{-7}$ | undetected | N/A | $1.50^{+0.13}_{-0.12}$ |
| EPIC 211068400 | Henyey | ⋯ | $-2.46 \pm 0.23$ | no data | ⋯ | ⋯ | $0.12^{+0.01}_{-0.03}$ |
| FM Tau | $4.3^{+0.7}_{-0.6}$ | $0.39^{+0.05}_{-0.08}$ | $-0.41 \pm 0.31$ | $0.7^{+0.5}_{-0.6}$ | $41.8^{+5.9}_{-5.9}$ | $3.7^{+0.9}_{-0.8}$ | $1.86^{+0.11}_{-0.11}$ |
| FP Tau | $1.3^{+0.6}_{-0.2}$ | $0.35^{+0.06}_{-0.04}$ | $-1.02 \pm 0.31$ | $0.2^{+0.6}_{-0.2}$ | $5.9^{+1.1}_{-0.9}$ | $0.55^{+0.25}_{-0.17}$ | $0.26^{+0.03}_{-0.02}$ |
| FX Tau A | $2.9^{+0.7}_{-0.6}$ | $0.52^{+0.04}_{-0.04}$ | $-0.89 \pm 0.27$ | $1.2^{+0.8}_{-0.7}$ | $9.2^{+2.9}_{-2.5}$ | $0.51^{+0.27}_{-0.20}$ | $0.69^{+0.04}_{-0.03}$ |
| GK Tau A | $1.5^{+0.6}_{-0.4}$ | $0.56^{+0.06}_{-0.05}$ | $-0.53 \pm 0.22$ | $0.6^{+0.3}_{-0.3}$ | $15.2^{+6.4}_{-6.1}$ | $1.1^{+0.9}_{-0.6}$ | $1.74^{+0.05}_{-0.05}$ |
| GM Aur | $3.3^{+0.8}_{-0.9}$ | $0.83^{+0.05}_{-0.07}$ | $-0.49 \pm 0.63$ | no data | ⋯ | ⋯ | $0.59^{+0.02}_{-0.03}$ |
| GO Tau | $3.7^{+0.6}_{-0.4}$ | $0.43^{+0.04}_{-0.03}$ | $-0.79 \pm 0.34$ | $1.6^{+0.6}_{-0.6}$ | $7.9^{+1.6}_{-1.4}$ | $0.44^{+0.15}_{-0.12}$ | $0.28^{+0.03}_{-0.02}$ |
| GSS 26 | $1.2^{+0.3}_{-0.2}$ | $0.42^{+0.07}_{-0.06}$ | $-0.17 \pm 0.19$ | no data | ⋯ | ⋯ | $0.53^{+0.03}_{-0.04}$ |
| GSS 37 W | $1.1^{+0.3}_{-0.2}$ | $0.50^{+0.03}_{-0.04}$ | $-0.83 \pm 0.19$ | no data | ⋯ | ⋯ | $0.47^{+0.03}_{-0.02}$ |
| GSS 39 | $0.9^{+0.2}_{-0.8}$ | $0.34^{+0.04}_{-0.04}$ | $-0.41 \pm 0.23$ | $15.1^{+4.4}_{-4.1}$ | $192^{+128}_{-74.7}$ | $55^{+65}_{-30}$ | $1.82^{+0.07}_{-0.07}$ |
| GV Tau S | $0.54^{+0.15}_{-0.10}$ | $0.64^{+0.06}_{-0.05}$ | $1.11 \pm 0.25$ | $10.1^{+4.9}_{-2.6}$ | $811^{+794}_{-322}$ | $234^{+413}_{-131}$ | $2.24^{+0.15}_{-0.18}$ |
| [GY92] 33 | $3.5^{+1.0}_{-1.0}$ | $0.71^{+0.07}_{-0.07}$ | $-0.22 \pm 0.43$ | $17.2^{+3.3}_{-3.2}$ | $27.8^{+30.7}_{-18.0}$ | $1.6^{+2.9}_{-1.2}$ | $0.29^{+0.02}_{-0.02}$ |
| [GY92] 235 | $2.1^{+1.2}_{-0.9}$ | $0.37^{+0.06}_{-0.12}$ | $0.02 \pm 0.34$ | no data | ⋯ | ⋯ | $2.21^{+0.18}_{-0.17}$ |
| Haro 1-16 | $3.3^{+1.5}_{-1.0}$ | $0.97^{+0.11}_{-0.15}$ | $-0.37 \pm 0.30$ | $0.2^{+1.7}_{-0.2}$ | $18.8^{+4.5}_{-1.4}$ | $0.9^{+0.4}_{-0.2}$ | $2.33^{+0.15}_{-0.17}$ |
| Haro 6-13 | $1.2^{+1.0}_{-0.5}$ | $0.60^{+0.08}_{-0.09}$ | $-0.02 \pm 0.30$ | $6.4^{+1.1}_{-1.2}$ | $169^{+41.4}_{-35.2}$ | $25^{+16}_{-11}$ | $1.55^{+0.12}_{-0.13}$ |
| Haro 6-33 | $1.9^{+0.8}_{-0.6}$ | $0.55^{+0.06}_{-0.04}$ | $0.09 \pm 0.26$ | $5.2^{+3.3}_{-1.7}$ | $8.3^{+6.2}_{-3.1}$ | $0.50^{+0.66}_{-0.26}$ | $0.55^{+0.03}_{-0.03}$ |
| Haro 6-28 | $4.1^{+0.6}_{-0.6}$ | $0.31^{+0.03}_{-0.03}$ | $-0.74 \pm 0.23$ | $1.9^{+0.8}_{-0.4}$ | $8.9^{+1.9}_{-1.3}$ | $0.6^{+0.2}_{-0.1}$ | $0.30^{+0.03}_{-0.03}$ |
| HIP 110526 E | Henyey | ⋯ | $-2.58 \pm 0.10$ | $0.3^{+0.2}_{-0.3}$ | undetected | N/A | $0.11^{+0.01}_{-0.02}$ |
| HIP 110526 W | Henyey | ⋯ | $-2.58 \pm 0.10$ | $0.1^{+0.4}_{-0.1}$ | undetected | N/A | $0.00^{+0.01}_{-0.00}$ |
| HIP 113579 | Henyey | ⋯ | $-2.72 \pm 0.10$ | $0.0^{+0.3}_{-0.0}$ | undetected | N/A | $0.01^{+0.06}_{-0.03}$ |
| HIP 11437NW | Henyey | ⋯ | $-2.55 \pm 0.19$ | no data | ⋯ | ⋯ | $0.08^{+0.01}_{-0.01}$ |
| HIP 11437SE | Henyey | ⋯ | $-2.55 \pm 0.19$ | no data | ⋯ | ⋯ | $-0.03^{+0.08}_{-0.02}$ |
| HIP 12545 | Henyey | ⋯ | $-2.66 \pm 0.10$ | no data | ⋯ | ⋯ | $-0.11^{+0.02}_{-0.01}$ |
| HIP 12635 | Henyey | ⋯ | $-2.59 \pm 0.14$ | no data | ⋯ | ⋯ | $-0.04^{+0.06}_{-0.05}$ |
| HIP 16563NW | Henyey | ⋯ | $-2.68 \pm 0.10$ | no data | ⋯ | ⋯ | $-0.04^{+0.07}_{-0.01}$ |
| HIP 16563SE | $53.1^{+3.7}_{-9.5}$ | $0.67^{+0.01}_{-0.01}$ | $-2.68 \pm 0.10$ | no data | ⋯ | ⋯ | $-0.01^{+0.00}_{-0.01}$ |
| HIP 17695 | Henyey | ⋯ | $-2.58 \pm 0.10$ | no data | ⋯ | ⋯ | $0.01^{+0.00}_{-0.01}$ |
| HIP 18859 | Henyey | ⋯ | $-2.70 \pm 0.17$ | no data | ⋯ | ⋯ | $0.37^{+0.05}_{-0.05}$ |
| HIP 81084 | Henyey | ⋯ | $-2.66 \pm 0.10$ | $0.0^{+0.6}_{-0.0}$ | $0.2^{+0.1}_{-0.1}$ | $0.0015^{+0.0012}_{-0.001}$ | $0.00^{+0.01}_{-0.00}$ |
| HIP 82688 | Henyey | ⋯ | $-2.75 \pm 0.10$ | no data | ⋯ | ⋯ | $0.07^{+0.02}_{-0.03}$ |
| HK Tau A | $1.8^{+0.5}_{-0.4}$ | $0.46^{+0.03}_{-0.04}$ | $-0.42 \pm 0.44$ | $2.8^{+1.2}_{-0.9}$ | $9.6^{+4.7}_{-3.6}$ | $0.70^{+0.59}_{-0.36}$ | $0.74^{+0.03}_{-0.03}$ |
| HO Tau | $18.0^{+4.9}_{-4.7}$ | $0.43^{+0.06}_{-0.03}$ | $-0.95 \pm 0.25$ | $0.2^{+0.8}_{-0.2}$ | $50.1^{+8.2}_{-4.3}$ | $2.6^{+0.9}_{-0.5}$ | $1.08^{+0.05}_{-0.05}$ |

**Table 2**
(Continued)

| Object | Age (Myr) | Mass ($M_\odot$) | Index | $A_V$ (mag) | Br$\gamma$ $\times$ $10^{-15}$ (erg s$^{-1}$ cm$^{-2}$) | $\dot{M}$ $\times$ $10^{-9}$ ($M_\odot$ yr$^{-1}$) | Veiling ($rK$) |
|---|---|---|---|---|---|---|---|
| HP Tau | $2.0^{+0.8}_{-0.6}$ | $1.10^{+0.05}_{-0.05}$ | $-0.43 \pm 0.26$ | $2.0^{+0.4}_{-0.2}$ | $93.3^{+18.1}_{-16.0}$ | $7.0^{+2.7}_{-2.1}$ | $1.80^{+0.09}_{-0.08}$ |
| IP Tau | $16.7^{+4.0}_{-2.5}$ | $0.69^{+0.03}_{-0.04}$ | $-0.88 \pm 0.25$ | $0.2^{+0.3}_{-0.2}$ | $8.6^{+3.5}_{-3.2}$ | $0.24^{+0.15}_{-0.12}$ | $1.24^{+0.05}_{-0.04}$ |
| IQ Tau | $1.6^{+0.6}_{-0.4}$ | $0.48^{+0.06}_{-0.04}$ | $-0.84 \pm 0.26$ | $2.4^{+1.2}_{-1.0}$ | $60.9^{+13.7}_{-10.6}$ | $7.1^{+3.3}_{-2.3}$ | $0.91^{+0.04}_{-0.04}$ |
| IRAS 03260+3111 B | $0.46^{+1.12}_{-0.33}$ | $0.24^{+0.04}_{-0.04}$ | … | … | … | … | $0.42^{+0.05}_{-0.06}$ |
| IRAS 03301+3111 | $1.1^{+0.2}_{-0.2}$ | $0.39^{+0.05}_{-0.05}$ | $0.38 \pm 0.17$ | $8.2^{+3.3}_{-1.7}$ | $338^{+174}_{-78}$ | $92^{+88}_{-36}$ | $1.27^{+0.08}_{-0.06}$ |
| IRAS 04108+2803 E | $1.8^{+0.9}_{-0.7}$ | $0.71^{+0.07}_{-0.09}$ | $1.16 \pm 0.25$ | $19^{+6}_{-5}$ | $36.9^{+38.0}_{-17.1}$ | $2.9^{+5.4}_{-1.8}$ | $0.89^{+0.05}_{-0.06}$ |
| IRAS 04113+2758 S | $0.31^{+0.66}_{-0.10}$ | $0.24^{+0.02}_{-0.01}$ | $-0.02 \pm 0.24$ | $7.6^{+2.9}_{-1.2}$ | $98^{+58}_{-29}$ | $32^{+29}_{-12}$ | $1.02^{+0.02}_{-0.03}$ |
| IRAS 04181+2665 M | $2.2^{+0.8}_{-0.6}$ | $0.48^{+0.05}_{-0.04}$ | $0.38 \pm 0.26$ | $37^{+2}_{-6}$ | undetected | N/A | $0.21^{+0.03}_{-0.03}$ |
| IRAS 04181+2665 S | $2.9^{+1.2}_{-1.5}$ | $0.38^{+0.07}_{-0.10}$ | $0.64 \pm 0.30$ | $24.6^{+6.4}_{-7.6}$ | $149^{+153}_{-84}$ | $22^{+48}_{-17}$ | $1.98^{+0.18}_{-0.19}$ |
| IRAS 04295+2251 | $1.3^{+1.3}_{-0.4}$ | $0.38^{+0.10}_{-0.07}$ | $0.66 \pm 0.18$ | $18^{+7}_{-4}$ | $67.6^{+77.4}_{-25.3}$ | $11^{+27}_{-6.9}$ | $1.98^{+0.17}_{-0.18}$ |
| IRAS 04489+3042 | $2.6^{+1.0}_{-1.4}$ | $0.34^{+0.05}_{-0.15}$ | $0.28 \pm 0.23$ | no data | … | … | $1.58^{+0.13}_{-0.10}$ |
| IRAS 04591−0856 | $1.2^{+0.3}_{-1.0}$ | $0.31^{+0.08}_{-0.05}$ | $0.51 \pm 0.29$ | $14^{+4.5}_{-2.5}$ | undetected | N/A | $0.71^{+0.10}_{-0.09}$ |
| IRAS 05379−0758(2) | $0.70^{+0.34}_{-0.58}$ | $0.37^{+0.06}_{-0.06}$ | $0.74 \pm 0.47$ | $9.2^{+3.8}_{-1.7}$ | $35.9^{+26.0}_{-11.3}$ | $6.8^{+8.9}_{-3.3}$ | $0.58^{+0.06}_{-0.06}$ |
| IRAS 05555−1405(4) | $0.47^{+1.7}_{-0.22}$ | $0.77^{+0.12}_{-0.13}$ | $0.45 \pm 0.33$ | $3.8^{+2.2}_{-1.3}$ | $7.8^{+9.6}_{-6.4}$ | $0.6^{+1.8}_{-0.6}$ | $0.87^{+0.43}_{-0.27}$ |
| IRAS 16285−2358 | $3.1^{+1.8}_{-1.0}$ | $0.34^{+0.07}_{-0.07}$ | $0.12 \pm 0.18$ | $8.6^{+8.4}_{-2.6}$ | $31.7^{+52.1}_{-11.0}$ | $3.2^{+9.5}_{-1.6}$ | $1.69^{+0.13}_{-0.12}$ |
| IRAS 16288−2450 W2 | $2.8^{+0.7}_{-0.6}$ | $0.46^{+0.02}_{-0.06}$ | $-0.15 \pm 0.18$ | $13.7^{+1.8}_{-4.7}$ | $109^{+38}_{-55}$ | $13^{+7.7}_{-8.1}$ | $0.41^{+0.03}_{-0.04}$ |
| IRAS 19247+2238(1) | $0.95^{+0.50}_{-0.19}$ | $0.53^{+0.06}_{-0.05}$ | $-0.13 \pm 0.39$ | $1.6^{+2.2}_{-0.8}$ | $335^{+127}_{-59}$ | $70^{+51}_{-23}$ | $2.00^{+0.08}_{-0.08}$ |
| IRAS 19247+2238(2) | $1.3^{+0.7}_{-0.3}$ | $0.49^{+0.08}_{-0.07}$ | $-0.13 \pm 0.39$ | $2.2^{+1.4}_{-0.7}$ | $41.4^{+26.2}_{-18.2}$ | $4.7^{+5.4}_{-2.8}$ | $1.41^{+0.06}_{-0.06}$ |
| LkCa 15 | $8.9^{+4.9}_{-4.8}$ | $1.09^{+0.03}_{-0.03}$ | $-1.00 \pm 0.31$ | $0.6^{+0.6}_{-0.6}$ | $85.4^{+27.9}_{-24.4}$ | $4.2^{+2.7}_{-1.9}$ | $1.07^{+0.05}_{-0.06}$ |
| [TS84] IRS 5 NE | $0.19^{+0.99}_{-0.07}$ | $0.28^{+0.05}_{-0.03}$ | $1.11 \pm 0.35$ | no data | … | … | $0.38^{+0.05}_{-0.04}$ |
| ROX 25 | $3.0^{+5.2}_{-1.2}$ | $1.42^{+0.05}_{-0.05}$ | $-0.63 \pm 0.16$ | no data | … | … | $1.05^{+0.19}_{-0.15}$ |
| ROX 27 | $1.3^{+0.5}_{-0.3}$ | $0.66^{+0.04}_{-0.05}$ | $-0.33 \pm 0.25$ | no data | … | … | $1.28^{+0.04}_{-0.05}$ |
| SR 24 S | $0.98^{+1.9}_{-0.59}$ | $0.89^{+0.11}_{-0.13}$ | $-0.30 \pm 0.30$ | $3.8^{+0.9}_{-0.8}$ | $141^{+20.0}_{-17.5}$ | $16^{+13}_{-9.5}$ | $3.03^{+0.53}_{-0.65}$ |
| T Tauri N | $0.71^{+0.36}_{-0.20}$ | $0.82^{+0.07}_{-0.08}$ | $0.08 \pm 0.40$ | $0.0^{+0.4}_{-0.0}$ | $1371^{+207}_{-145}$ | $335^{+130}_{-91}$ | $3.00^{+0.04}_{-0.04}$ |
| TWA 3 B | $7.4^{+2.1}_{-0.9}$ | $0.33^{+0.03}_{-0.01}$ | $-0.69 \pm 0.35$ | $0.12^{+0.58}_{-0.12}$ | $6.6^{+3.4}_{-2.9}$ | … | $-0.25^{+0.01}_{-0.02}$ |
| TWA 4 | $16.5^{+15.7}_{-1.4}$ | $0.79^{+0.09}_{-0.01}$ | $-0.60 \pm 0.51$ | no data | … | … | $0.12^{+0.01}_{-0.11}$ |
| TWA 6 A | $15.2^{+2.5}_{-1.4}$ | $0.90^{+0.01}_{-0.01}$ | $-2.60 \pm 0.10$ | $0.0^{+0.2}_{-0.0}$ | undetected | … | $-0.17^{+0.06}_{-0.01}$ |
| TWA 7 | $42.0^{+3.6}_{-3.3}$ | $0.50^{+0.02}_{-0.01}$ | $-2.44 \pm 0.17$ | $0.5^{+0.4}_{-0.2}$ | undetected | N/A | $-0.05^{+0.02}_{-0.02}$ |
| TWA 8 A | Henyey | … | $-2.61 \pm 0.10$ | $0.5^{+0.3}_{-0.3}$ | $6.1^{+2.1}_{-2.0}$ | $0.08^{+0.04}_{-0.03}$ | $-0.08^{+0.02}_{-0.03}$ |
| TWA 9 A | $14.0^{+4.6}_{-2.3}$ | $0.85^{+0.03}_{-0.03}$ | $-2.59 \pm 0.12$ | $0.0^{+0.2}_{-0.0}$ | undetected | N/A | $-0.01^{+0.04}_{-0.04}$ |
| TWA 9 B | $12.1^{+2.8}_{-1.5}$ | $0.49^{+0.04}_{-0.03}$ | $-2.59 \pm 0.12$ | $0.0^{+0.9}_{-0.0}$ | undetected | N/A | $-0.05^{+0.11}_{-0.13}$ |
| TWA 13 N | $11.0^{+2.3}_{-1.8}$ | $0.66^{+0.01}_{-0.02}$ | $-2.53 \pm 0.17$ | $0.5^{+0.5}_{-0.3}$ | undetected | N/A | $-0.13^{+0.11}_{-0.01}$ |
| TWA 13 S | $8.4^{+1.0}_{-0.9}$ | $0.66^{+0.01}_{-0.01}$ | $-2.77 \pm 0.08$ | $0.6^{+0.6}_{-0.4}$ | undetected | N/A | $-0.11^{+0.02}_{-0.01}$ |
| TWA 23 | $12.8^{+2.5}_{-2.8}$ | $0.52^{+0.02}_{-0.02}$ | $-2.57 \pm 0.10$ | no data | … | … | $-0.04^{+0.02}_{-0.05}$ |
| TYC 915-1391-1 | $1.4^{+0.2}_{-0.1}$ | $0.37^{+0.02}_{-0.01}$ | $-2.60 \pm 0.11$ | $0.4^{+0.9}_{-0.4}$ | undetected | N/A | $-0.16^{+0.02}_{-0.01}$ |
| TYC 6349-0200-1 NW | $19.3^{+17.0}_{-0.8}$ | $0.86^{+0.04}_{-0.02}$ | $-2.70 \pm 0.09$ | no data | … | … | $0.08^{+0.02}_{-0.06}$ |
| TYC 6349-0200-1 SE | Henyey | … | $-2.70 \pm 0.09$ | no data | … | … | $-0.22^{+0.02}_{-0.06}$ |
| TYC 6878-0195-1 | $19.5^{+4.9}_{-4.8}$ | $0.82^{+0.04}_{-0.02}$ | $-2.55 \pm 0.10$ | no data | … | … | $0.07^{+0.02}_{-0.09}$ |
| UY Aur NE | $1.0^{+0.2}_{-0.2}$ | $0.50^{+0.04}_{-0.04}$ | $-0.01 \pm 0.16$ | no data | … | … | $1.23^{+0.05}_{-0.05}$ |
| V347 Aur | $>0.1^{+0.73}_{-?}$ | $0.25^{+0.07}_{-0.07}$ | $-0.10 \pm 0.27$ | $1.3^{+1.0}_{-0.6}$ | $426^{+80}_{-56}$ | $264^{+136}_{-92}$ | $1.17^{+0.09}_{-0.08}$ |
| V710 Tau N | $2.6^{+0.3}_{-0.3}$ | $0.46^{+0.01}_{-0.03}$ | $-0.91 \pm 0.23$ | $1.6^{+0.7}_{-0.7}$ | $33.8^{+6.7}_{-5.8}$ | $3.0^{+1.0}_{-0.8}$ | $0.27^{+0.02}_{-0.02}$ |
| V710 Tau S | $2.9^{+0.3}_{-0.2}$ | $0.40^{+0.03}_{-0.02}$ | $-0.91 \pm 0.23$ | $1.3^{+0.5}_{-0.5}$ | undetected | N/A | $0.22^{+0.01}_{-0.02}$ |
| WSB 82 | $1.5^{+0.6}_{-0.4}$ | $0.81^{+0.13}_{-0.18}$ | $-0.51 \pm 0.19$ | no data | … | … | $4.15^{+0.44}_{-0.70}$ |
| WL 20 E | $1.7^{+0.7}_{-0.5}$ | $0.55^{+0.07}_{-0.04}$ | $0.85 \pm 0.38$ | no data | … | … | $0.32^{+0.03}_{-0.03}$ |
| WL 20 W | $3.1^{+1.1}_{-0.8}$ | $0.39^{+0.04}_{-0.04}$ | $0.85 \pm 0.38$ | no data | … | … | $0.14^{+0.03}_{-0.02}$ |
| WLY 2-42 | $1.6^{+0.6}_{-0.1}$ | $0.31^{+0.04}_{-0.03}$ | $0.10 \pm 0.28$ | no data | … | … | $0.10^{+0.02}_{-0.03}$ |
| YLW 58 | $0.24^{+2.6}_{-0.13}$ | $0.16^{+0.06}_{-0.02}$ | $-0.54 \pm 0.29$ | $1.9^{+1.5}_{-0.9}$ | $17.0^{+5.6}_{-3.8}$ | $4.2^{+3.3}_{-1.9}$ | $1.68^{+0.11}_{-0.15}$ |

### 3.2. *Extinction*

To estimate the extinction to each target, we converted the $K$-band photospheric temperature, as determined from high-resolution spectroscopy, to a spectral type using Table 5 in G. J. Herczeg & L. A. Hillenbrand (2014). We then selected the matching spectrum from the SpeX library (J. T. Rayner et al. 2009) to use as the template. We then used the procedure described in M. S. Connelley & T. P. Greene (2010) to simultaneously fit the library template spectrum to the observed YSO spectrum by varying the extinction and the veiling temperature. We held the veiling constant at the value determined by the high-resolution spectral analysis. Objects that were not observed by SpeX are shown in Table 2 as "no data."

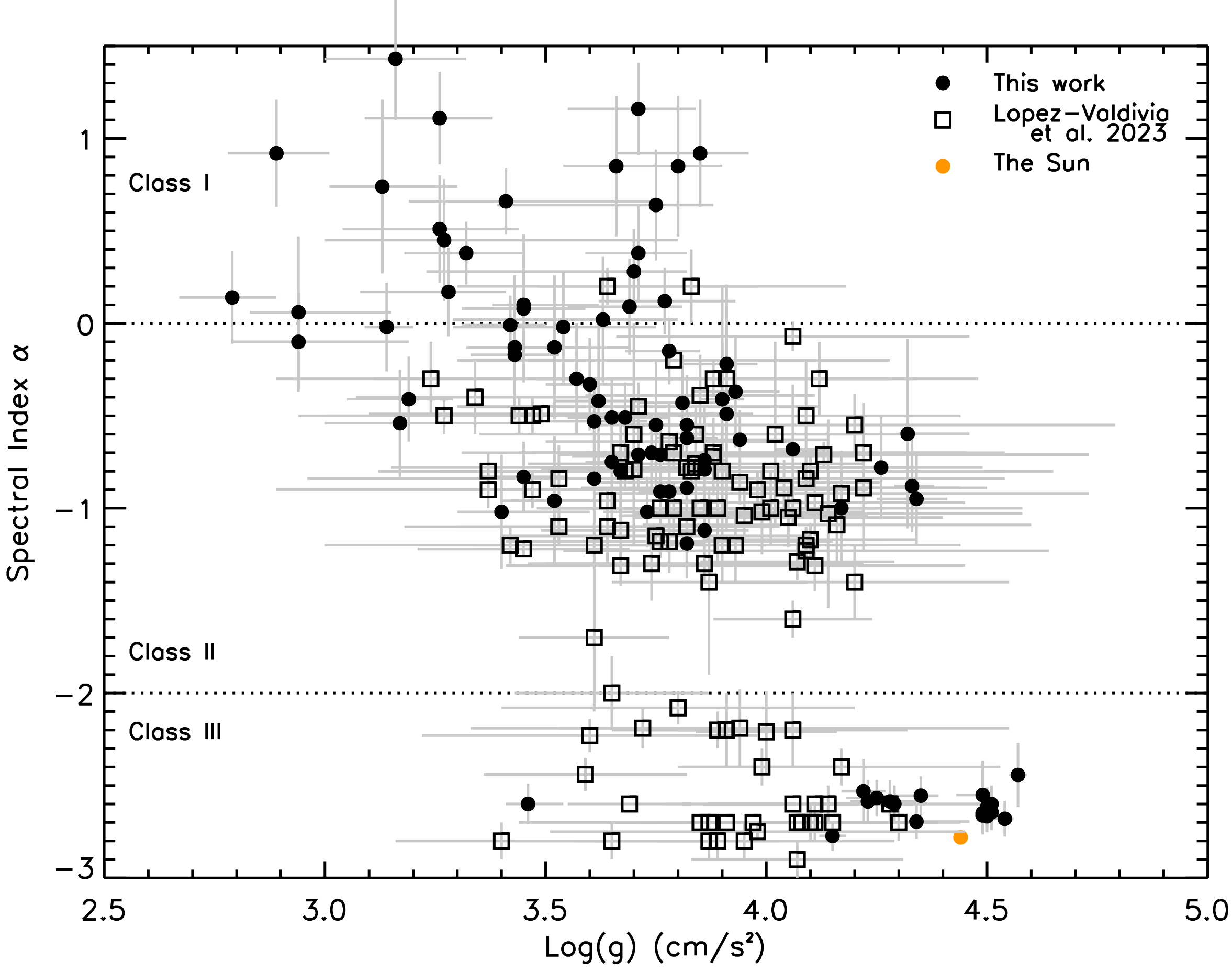

**Figure 1.** Along with a broad trend toward lower spectral index with increasing gravity (age), there is an $\sim\pm1$ spread in the spectral index at any given log($g$). While the classes are often assumed to be an evolutionary sequence, there are Class I, II, and III objects of the same age, showing that the amount of circumstellar material varies widely at any given time. The dearth of targets between the Class II and III objects is likely a selection effect, because we chose Class II YSOs with massive disks.

### *3.3. Spectral Index*

We calculated the spectral index from 2 to 22 $\mu$m using data from the ALLWISE catalog. The spectral index is defined as $d\log(\lambda F_\lambda)/d\log(\lambda)$ (C. J. Lada 1987). For each star, we found the spectral index as the best-fit linear slope between $\log(\lambda F_\lambda)$ and $\log(\lambda)$ over the specified wavelength range, along with the uncertainty in the linear slope.

The spectral index can be affected by high extinction, primarily among the Class I YSOs. We used the extinction law from R. Indebetouw et al. (2005) to deredden the 2–22 $\mu$m IR fluxes, then we recalculated the spectral index as described above. The corrected spectral index can be estimated by $\alpha_{\rm corrected} = \alpha - 0.027A_V$.

### *3.4. Brγ Flux and Accretion Rate*

The Br$\gamma$ line flux was measured in IRAF using the SpeXtool pseudoflux-calibrated spectrum, where the spectrum is calibrated relative to the spectrum of the telluric standard star, which has a known magnitude. SpeX data were used because early versions of iSHELL's data reduction code did not preserve hydrogen line fluxes through a telluric correction. Similar to extinction, objects not observed by SpeX are noted in Table 2 as "no data." When an object was observed, but no Br$\gamma$ emission was detected, we put "undetected" in Table 2. We further used a relation between the pseudoflux-calibrated continuum and the 2MASS $K$-band magnitude of the star to scale the line flux, in case of an error in the SpeXtool calibration due to (for example) nonphotometric observing conditions. The line flux of each source was then corrected for extinction (see Section 3.2) and normalized to a distance of 140 pc. We used Equation (2) from J. Muzerolle et al. (1998) to calculate the accretion luminosity, and then the relation $\dot{M} = (L_{\rm acc}R_*)/(GM_*)$ from L. Hartmann (1998) to calculate the mass accretion rate. $M_*$ was derived from the G. A. Feiden (2016) models and mostly depends on temperature, whereas $R_*$ is model dependent and is derived from the mass and gravity. The derivations of temperature, mass, and gravity are from the high-resolution spectra. The extraction of these properties from high-resolution $K$-band spectra is described in C. Flores et al. (2019), and the source of this information is noted in Table 1.

## 4. Results

### *4.1. Evolution of the Spectral Index*

Figure 1 shows the relationship between our calculated spectral index values and the gravities for the objects in our sample. We also include the Sun as the presumed end point of the evolutionary paths. Class II YSOs from R. López-Valdivia et al. (2023) help to fill in the gap in spectral index from $-1$ to $-2.5$. The evolutionary path for a star is expected to broadly proceed from the top left to lower right in this diagram, and we see a trend that is generally consistent with that expectation. Figure 2 (top) is similar to Figure 1, but the spectral index has been corrected for extinction. Here we see that several low-veiling Class I objects, after correcting for extinction, fall among the Class II YSOs.

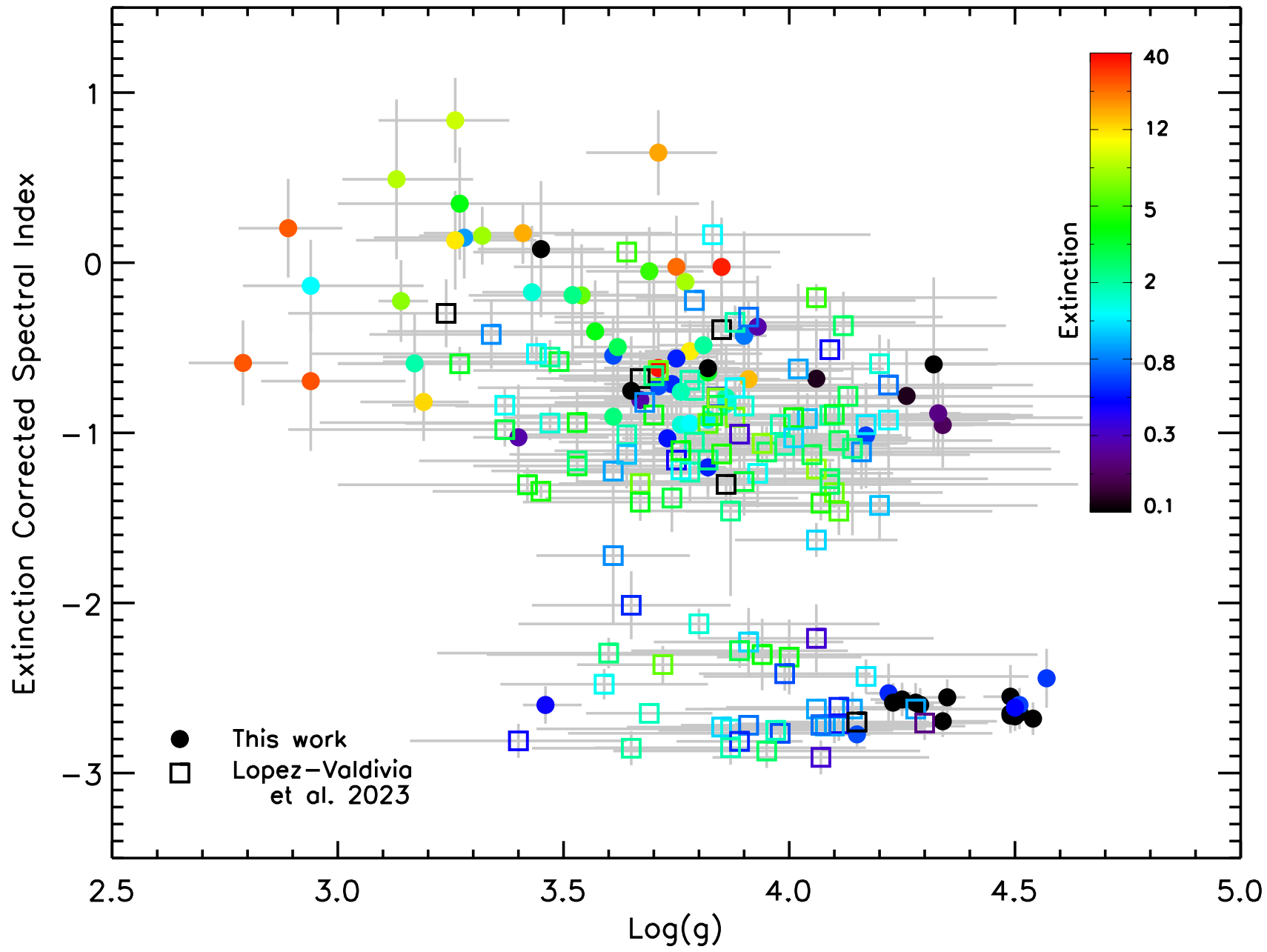

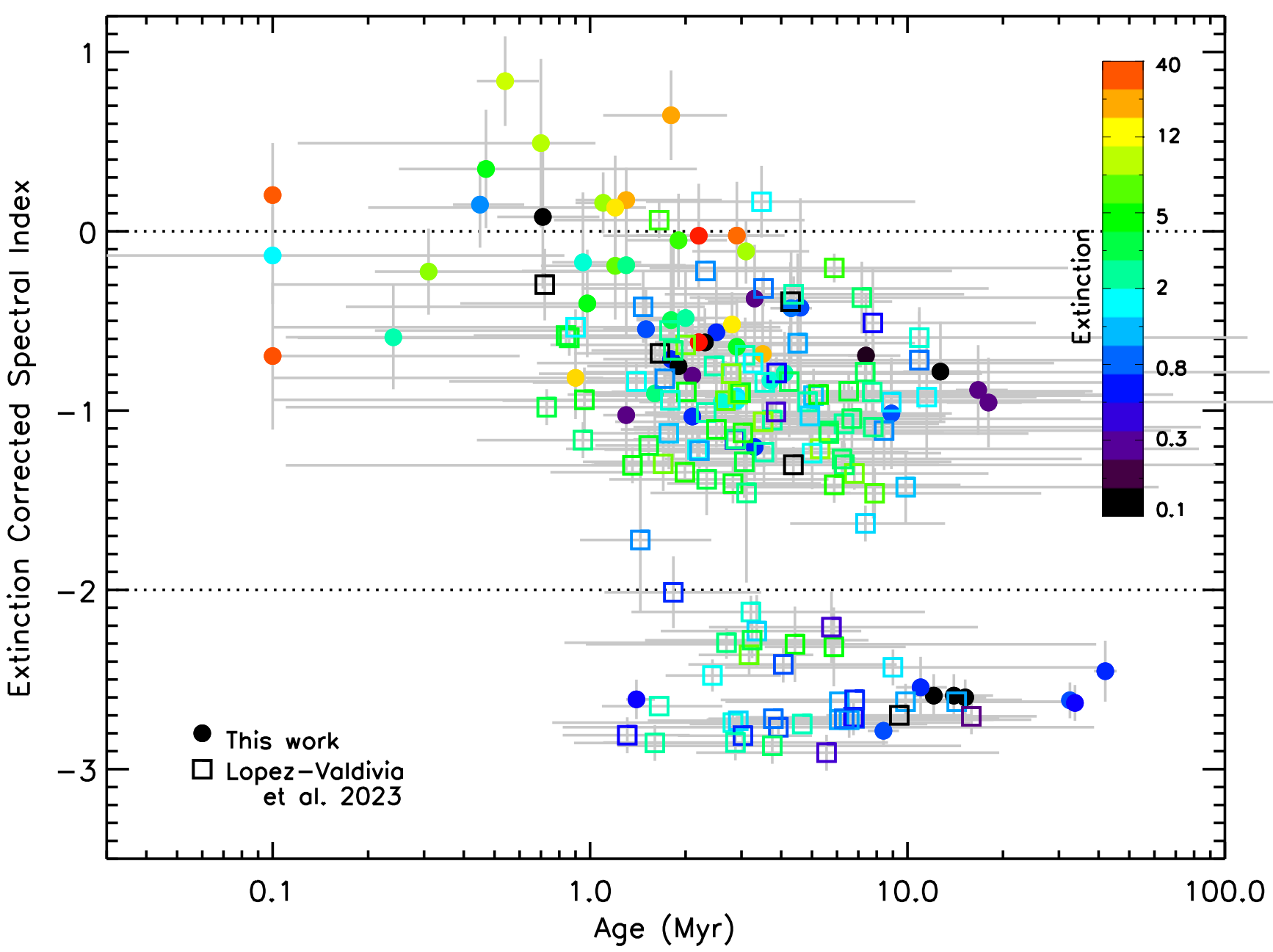

**Figure 2.** Top: after correcting the spectral index for extinction, many of the older Class I YSOs start to overlap with the Class II YSOs. These objects were higher in Figure 1 but moved down among the T Tauri stars in this figure. Bottom: plotting age (based on G. A. Feiden 2016 models) vs. extinction-corrected spectral index. A few stars have low gravity and are expected to be ∼1 Myr despite being optically visible Class III YSOs. Many other Class II YSOs are optically visible despite being less than 1 Myr old. We do not estimate ages beyond the limit of the G. A. Feiden (2016) models, or for stars hotter than 5000 K where the isochrones are very close to each other.

The figures establish that at any given gravity (and age), YSOs are found with a wide range of spectral indices. Taking $\log(g) = 3.7$ as an example (the gravity of many of our YSOs are near this value), we find that YSOs can have spectral indices from $+1.1$ to $-2.8$, and thus Class I, II, and III YSOs can all have the same age. The scatter in $\log(g)$ is not related to mass because there is only a slight dependence of $\log(g)$ on temperature.

While the spectral index of individual stars, and YSOs as a whole, decreases with time (Figure 1), stars may take different paths across the $\alpha$ versus log($g$) diagram from a deeply embedded protostar toward an optically revealed pre-main-sequence star. The data in Figure 2, where the spectral index has been corrected for extinction, follow a broad swath rather than a tight correlation, and this width is very important. One potential reason for the trend being this wide is that the dissipation of warm circumstellar material may be a highly stochastic process. The spectral indices of YSOs may evolve gradually at different rates, or ejections may rapidly change an object's spectral index. As further described in Sections 5.1, 5.4, and 5.6, extinction can affect the spectral index of YSOs, and there are a few Class I YSOs that are candidates for being T Tauri stars seen through high extinction, highlighting the problematic nature of using spectral index as a proxy for age.

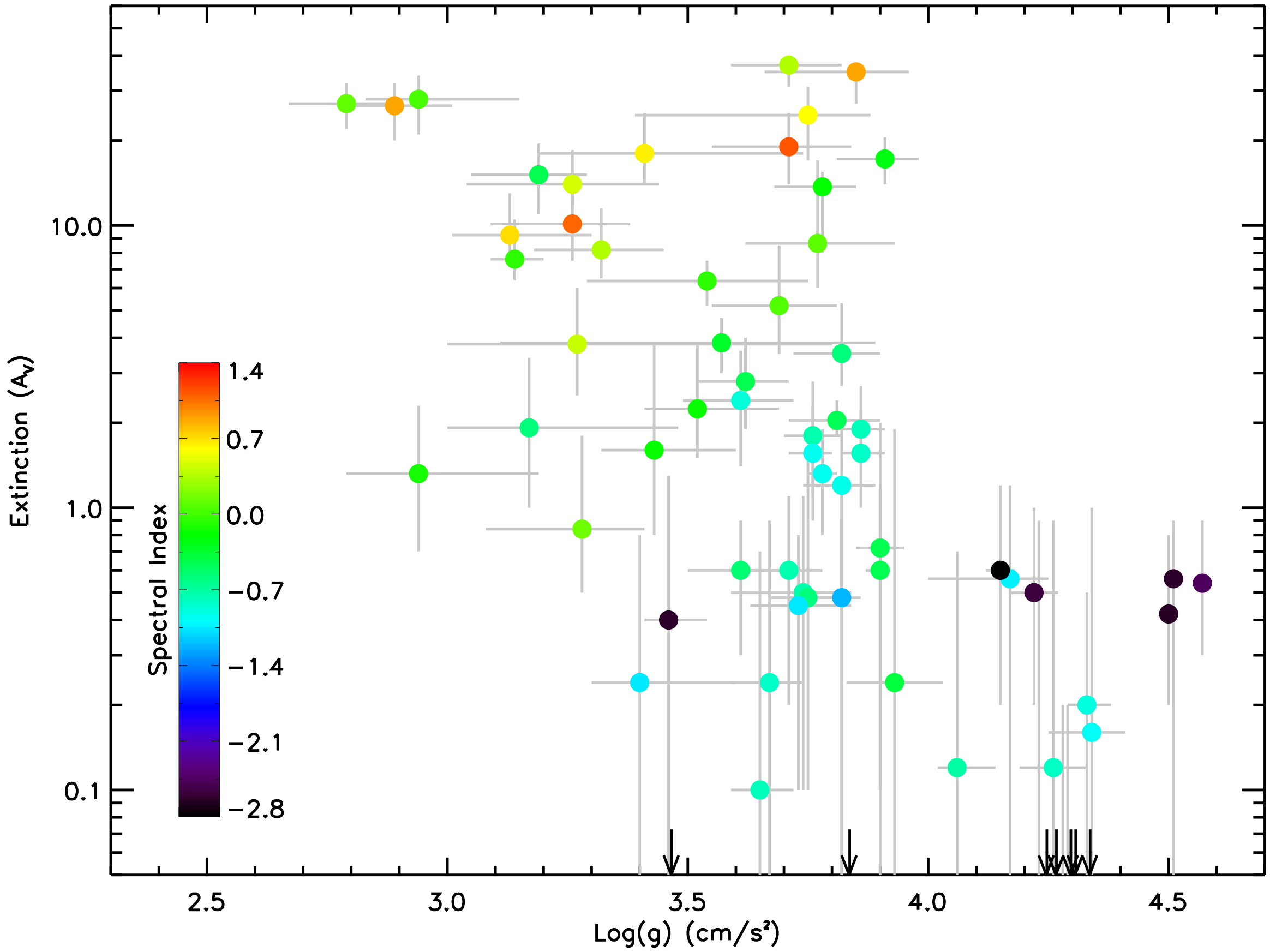

**Figure 3.** We see a wide range of extinction values for ages under ∼5 Myr (log($g$) < 4.0) with no apparent trend with time, suggesting that the process of revealing a star is very stochastic. Stars with similar ages have extinctions that range from nearly zero to ∼40 mag. After ∼5 Myr, no star has more than 1 mag of extinction. The color bar shows the spectral index (uncorrected for extinction).

The wide vertical spread of the data in Figure 1 shows that when the log($g$) of YSOs reaches ∼3, they already have a wide range of spectral index values, otherwise the targets in the upper left corner of Figure 1 would be more tightly clustered. The data are inconsistent with YSOs gradually dissipating their circumstellar material at similar rates from similar initial conditions. YSOs may originate on this plot with a range of spectral index values due to individually specific initial conditions and evolution before the Class I phase.

### 4.2. Extinction

The star formation process begins with a deeply embedded protostar that is eventually optically revealed as a young pre-main-sequence star. However, that process of dissipating the source of optical extinction may be slow and gradual, or sudden and rapid. Dynamical ejections are believed to be common (B. Reipurth 2000; B. Reipurth et al. 2010), in which case stochastic processes may play an important role in making a young star optically visible.

As expected, our sample of low-veiling Class I YSOs on average has higher extinctions than Class II or III YSOs. Class II YSOs in general are mostly found to have $A_V < 5$, although some Class II YSOs are seen through as much as 13 mag of visual extinction. If we had Class II YSOs with smaller disks in our sample, we may have had more objects with $A_V < 0.5$.

Figure 3 shows that at any given age, as traced by log($g$), YSOs can have a very wide range of extinctions along the line of sight. For example, at log($g$) ∼ 3.7, the extinction ranges from $A_V < 0.1$ to $A_V = 40$. We also note that instead of seeing a steady trend toward lower extinction with time, there appears to be an abrupt change toward low extinction near log($g$) = 4. These two findings, a wide spread in extinction at a given age and lack of a clear trend with age, both suggest that the process of optically revealing a YSO is highly stochastic, and is not a slow, steady, and gradual process. Young stars do need to start as deeply embedded objects and evolve to become optically visible main-sequence stars. However, because a young star being optically visible is not grounds alone to conclude that it is younger than a YSO seen through high extinction. Indeed, Figure 3 shows that there are many objects with low extinction at a younger age and objects that are older with higher extinction. Both deeply embedded and optically visible stars can be the same age, but at different stages of their evolution,

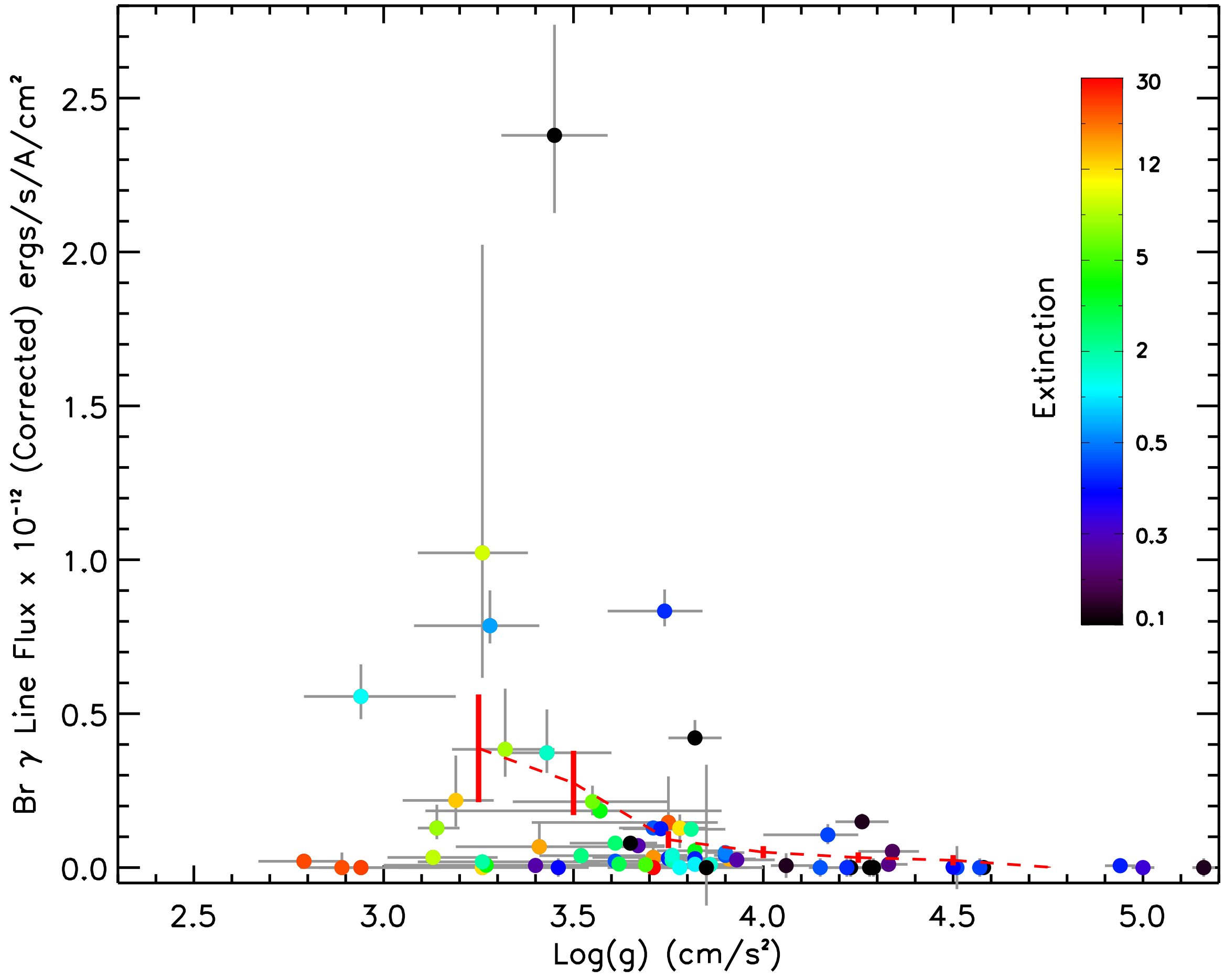

**Figure 4.** The Br$\gamma$ line flux, corrected for extinction and distance, overall declines as stars evolve. Two T Tauri stars are among the highest accretors, demonstrating that Class II YSOs can also have high accretion rates. The red dashed line shows the mean Br$\gamma$ line flux in $\log(g) = 0.5$ wide bins, with each bin separated by one-half this width, as well as the uncertainty in the mean of each bin. We find an overall trend of decreasing Br$\gamma$ line flux as the stars age, suggesting that the mass accretion rate does decrease with time on average, but there is a wide range of values at a given age. The color scale bar is extinction ($A_V$).

and an optically visible YSO can be younger than another that is still deeply embedded.

### *4.3. Mass Accretion Rate*

Mass accretion, often traced using hydrogen emission lines, is a fundamental observational characteristic of young stars. The presence of hydrogen emission lines was one of the properties that A. H. Joy (1945) used to define T Tauri stars as a class. J. Muzerolle et al. (1998) showed that the log of the Br$\gamma$ line luminosity is linearly proportional to the log of the accretion luminosity, establishing the Br$\gamma$ line as a useful mass accretion tracer in the near-IR. We also use the Br$\gamma$ line flux as a tracer of the mass accretion rate.

Low-veiling Class I YSOs and Class II YSOs have the same overall range of Br$\gamma$ line fluxes, from nearly zero up to $\sim 10^{-12}$ erg s$^{-1}$ A$^{-1}$ cm$^{-2}$. While low-veiling Class I objects have a higher average Br$\gamma$ line flux than Class II YSOs, a two-sample Kolmogorov–Smirnov test shows that the Br$\gamma$ line flux distributions (corrected for distance and extinction) for low-veiling Class I YSOs and Class II YSOs do not originate in different parent populations to 90% confidence. We find that the low-veiling Class I and Class II Br$\gamma$ line flux distributions are not statistically distinguishable. In contrast, recent results by E. Fiorellino et al. (2021, 2023) find that Class I and flat-spectrum YSOs have higher mass accretion rates than Class II YSOs. This discrepancy is likely due to our sample being limited to low-veiling Class I YSOs, and since veiling is connected to accretion, these low-veiling Class I YSOs are expected to have lower mass accretion rates than high-veiling Class I YSOs.

The Br$\gamma$ line flux, combining the low-veiling Class I, II, and III YSOs, decreases with increasing $\log(g)$, and hence age (see Figure 4). We took the average of the Br$\gamma$ line fluxes within $\log(g) = 0.5$ wide bins, with each successive bin offset by one-half of that width. We note that some bins, such as the lowest $\log(g)$ from 3.0 to 3.5, are limited by the small numbers of objects within that bin. We caution that this finding does not necessarily mean that the mass accretion rate gradually declines for a given YSO. Some of the mass of young stars may be gained in short-lived eruptions (e.g., EXors and FUors), which are not represented here. L. Hartmann et al. (1998) also found that the mass accretion rate for T Tauri stars declines with age, but with significant scatter that limited their constraint on the rate at which mass accretion declines.

Converting the Br$\gamma$ line flux to a mass accretion rate (Figure 5), we see that the mass accretion rates cover a wide range at a given age, have a weak dependence on age, and have a low dependence on spectral index. Figure 5 shows that the mean mass accretion rate (solid line in upper panel) does decrease with increasing $\log(g)$ and age but with an order of magnitude of scatter at a given age. Also, objects with a wide range of spectral indices are well mixed. The most notable trend with age is that the upper threshold of mass accretion rate appears to decline as gravity and age increase, approximately as $\dot{M}_{\rm max} \propto g^{-2}$.

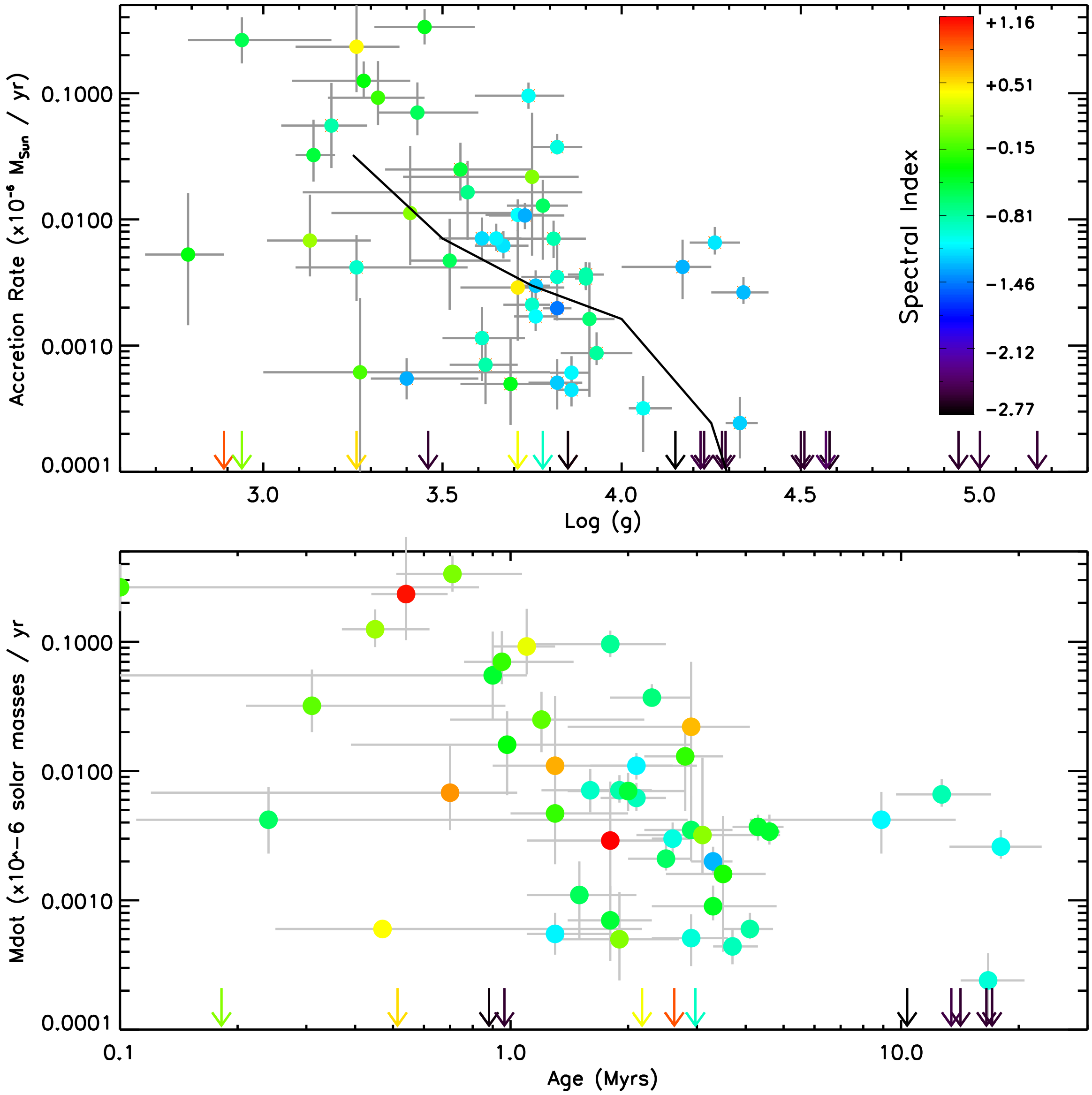

**Figure 5.** The mass accretion rate is shown vs. the log($g$) of the central star (top panel) and vs. the inferred age (lower panel) on a linear scale based on the log($g$), temperature, and G. A. Feiden (2016) models. The upper limit on the mass accretion rate decreases with time, but at any given age, the mass accretion rate can have a wide range of values, including zero. Class I and II YSOs, covering a wide range of spectra index values, are well mixed together. While a few YSOs with gravity log($g$) > 4.0 (>5 Myr of age) have detectable mass accretion, most objects older than this have negligible mass accretion rates.

There have been many previous efforts to understand the dependencies of the mass accretion rate on other stellar properties. C. F. Manara et al. (2020) looked at the dependence of $\dot{M}$ on disk mass, and did find a correlation with a scatter of about $\pm 1$ dex. C. F. Manara et al. (2021) considered the dependence of mass loss on stellar mass, finding a weak relationship between these parameters, with a scatter of about $\pm 1$ dex. J. M. Alcalá et al. (2017) showed that there is a correlation between $L_{\rm acc}$ and $L_*$ for a large sample of T Tauri stars, with the usual scatter of $\pm 1$ dex. Taken together, these studies show that the mass accretion rate has dependencies on many parameters (age, stellar mass, disk mass, and stellar luminosity), and with a scatter of about $\pm 1$ dex in each case. The large scatter in any one plot could be because $\dot{M}$ depends on many other parameters, in which case we see a lot of scatter when projecting a multidimensional relationship onto a 2D figure, or because the mass accretion rate is highly variable (e.g., M. S. Connelley & T. P. Greene 2014), or both.

Many objects, representing a wide range of ages and spectral indices, have low mass accretion rates below $10^{-9}\ M_\odot\ {\rm yr}^{-1}$. Nearly all Class III YSOs have accretion rates below this threshold. Figure 5 also shows that mass accretion, for the most part, drops below this threshold by the time log($g$) reaches four, approximately corresponding to an age of ~5 Myr. Magnetospheric mass accretion largely ends by the time stars are 5 Myr old.

### 4.4. Veiling

Veiling is another effect of mass accretion, where the photospheric absorption lines appear weaker due to continuum emission. As a mass accretion tracer, it was expected that the veiling should be higher for younger and more deeply embedded YSOs. We have measured the veiling in $K$ band as a byproduct of the modeling of the iSHELL data to retrieve the physical parameters of the stars (C. Flores et al. 2019), where $r_K$ is the $K$-band circumstellar continuum emission divided by the $K$-band stellar emission. A discussion of the evolution of veiling versus log($g$) is in Connelley et al. (2026).

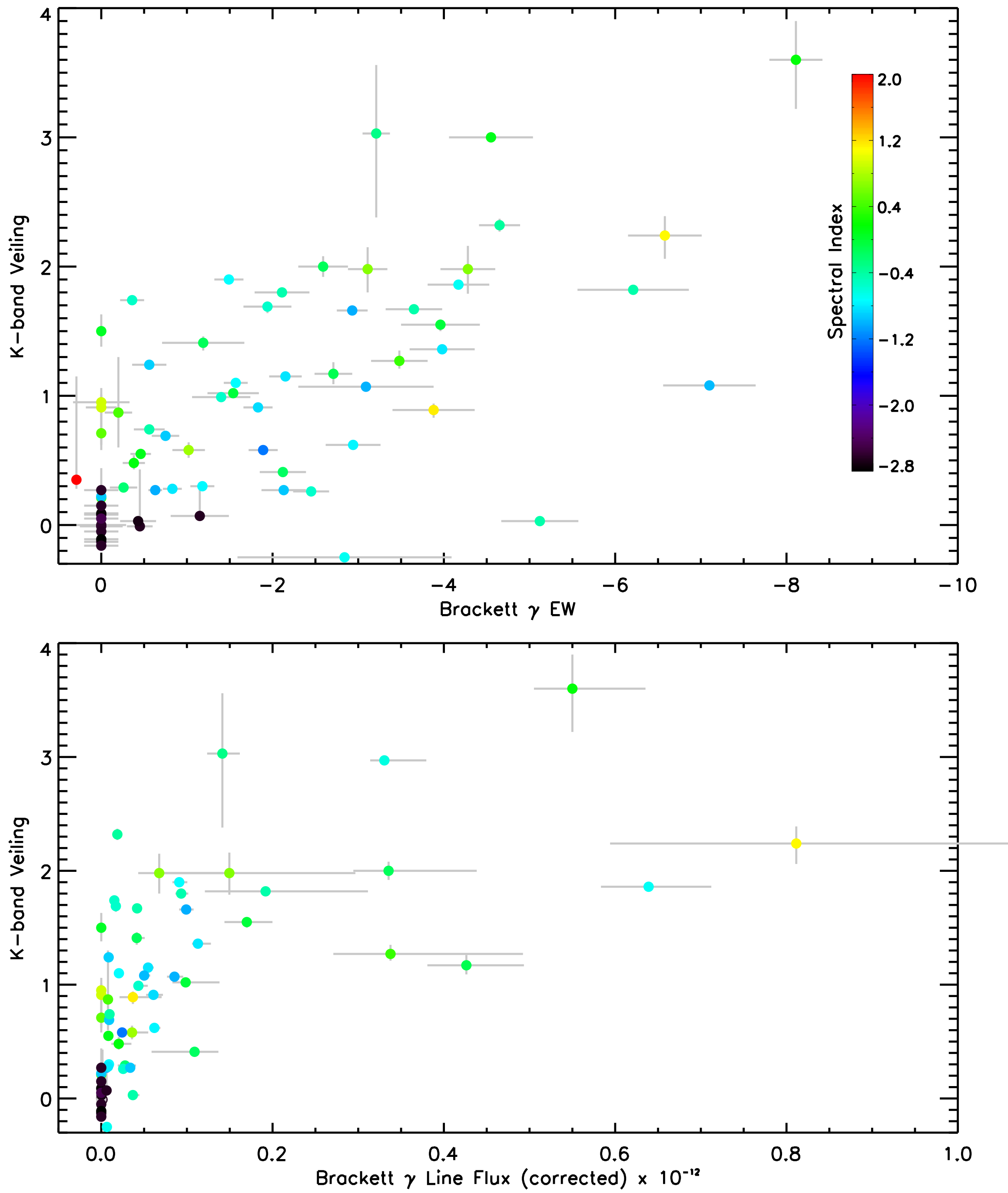

**Figure 6.** Here we compare veiling and the Br$\gamma$ line, as we use both as accretion tracers. The veiling and EW are measurements that are both relative to the continuum flux, and are loosely correlated. However, there is a poor connection between veiling and the Br$\gamma$ line flux (lower panel). It appears to show that targets can have some veiling before they start to show significant Br$\gamma$ flux in emission. All objects with a high Br$\gamma$ line flux also have high veiling, but not vice versa. There may be a warm nonaccreting inner disk contributing to the $K$-band continuum. The color bar reflects the spectral index, which is also accretion related.

As we are using veiling and the Br$\gamma$ line as accretion tracers, we expect them to positively correlate. Figure 6 (top) shows that the Br$\gamma$ EW does correlate with veiling, but with significant scatter. Many of the low-accreting objects still have substantial veiling, consistent with a warm nonaccreting inner disk. Figure 6 (bottom) also shows that objects with a low Br$\gamma$ line flux can have a wide range of veiling values. The similarity of the low-veiling Class I and Class II veiling distributions may indicate that their inner disks similarly contribute to veiling when the accretion is low.

## 5. Discussion

### 5.1. Are Class I Young Stellar Objects Younger Than Class II Young Stellar Objects?

A number of arguments has lead to the common belief that Class I YSOs are younger than Class II YSOs. Among other things, Class I YSOs have greater IR excess emission and high extinction, both of which are expected to decline as a star evolves. However, Figures 1 and 2 show that, consistent with C. Flores et al. (2024), the log($g$) distributions of low-veiling Class I and of Class II YSOs have significant overlap. However, C. Flores et al. (2024) have shown that the log($g$) distributions of low-veiling Class I and of Class II YSOs have only a small chance of having the same parent population, thus their log($g$) distributions are inconsistent with each other. It is important to note that both the low-veiling Class I and the Class II YSOs span a wide range in log($g$), implying a wide range of ages, from <1 to ∼20 Myr (using the G. A. Feiden 2016 models).

We stress that not all low-veiling Class I YSOs are younger than Class II YSOs, and that Class I YSOs are not a homogeneous group. We note that several Class II YSOs have log($g$) and temperature values consistent with being much younger than 1 Myr, and several low-veiling Class I YSOs have properties consistent with ages of ∼1 Myr or more. Furthermore, at any given log($g$), we see a wide range of spectral indices, from about −2 to +1 for log($g$) ∼ 3.7. This indicates that the observed flux at a given age can be dominated by either circumstellar emission or by the stellar photosphere.

In Figure 1, there is a group of low-veiling Class I YSOs with $\alpha \sim 1$ and log($g$) ∼ 3.7, which, after the spectral index is corrected for extinction, is found among the Class II objects. These low-veiling Class I YSOs are of similar age as Class II YSOs, as shown by their log($g$) values, but their high extinction makes them appear to be Class I. Thus, low-veiling Class I YSOs appear to be a combination of young (<1 Myr) embedded protostars and highly extincted Class II-like YSOs.

It should be emphasized that our results do not depend on luminosity. Many prior results infer ages based on the location of the stars on the H-R diagram, in comparison with theoretical evolutionary models. Stellar luminosities are very difficult to accurately measure, especially for Class I YSOs (G. W. Doppmann et al. 2005). The observed luminosity is affected by accretion, outflows, scattering, extinction, and IR disk emission. The impact of each of these is difficult to accurately quantify and correct for. Furthermore, theoretical models predict luminosity based on an effective temperature. C. Flores et al. (2022) found that the best-fit spectroscopic temperature for T Tauri stars depends on the observational wavelength (temperatures measured in the optical tend to be higher than in the near-IR), making it difficult to infer the effective temperature from an observed temperature and thus accurately predict the luminosity from the models. Each of these factors frustrates efforts to infer the relative ages of Class I and II YSOs from their observed luminosities, and we have therefore avoided using luminosity in our analysis.

### *5.2. Very Young Optically Visible Young Stellar Objects*

There is a number of optically visible YSOs that are among the youngest stars in our sample. When compared against the G. A. Feiden (2016) magnetic models, their gravities and temperatures suggest that they can be less than 1 Myr old. These optically visible stars present an opportunity to apply optical techniques and diagnostics to objects so young that they, in many cases, would be expected to be embedded.

Two well known examples are T Tauri N and V347 Aur. Both stars are optically visible and bright in the IR, and both have low gravities (log($g$) < 3.5), suggesting an age younger than 1 Myr. Other examples include IRAS 05555-1405(4) in a small optically visible group, which has a gravity of log($g$) = 3.2, suggesting an age less than 0.5 Myr. IRAS 19247 +2238 is a visible binary with an inferred age of ∼1 Myr. Among the Class II YSOs, FP Tau and UY Aur (NE) both have surface gravities of log($g$) = 3.4, suggesting they are both less than 1 Myr old.

Among the Class III YSOs, TYC 915-1391-1, V1023 Tau, and V1095 Tau are unusually young. These three stars have very low spectral indices, are optically visible, and have gravities suggesting ages of 1 Myr or less. This suggests that they were created by an unusual event; we speculate that they are orphaned protostars that were ejected from a multiple Class I or II system as discussed by B. Reipurth et al. (2010).

### *5.3. The Rate of Mass Growth*

R. J. White & L. A. Hillenbrand (2004) found no difference in the mass accretion rates between Class I and Class II YSOs, which is problematic as the (presumably) younger Class I YSOs should be accreting at a higher rate, which is broadly consistent with our findings. Taken at face value, Figure 7 suggests that most young stars do not experience significant mass growth after 1 Myr of age, and possibly that there is often insignificant mass growth after 0.1 Myr of age. As such, YSOs are likely to be very near their final mass when they become detectable in the near-IR as Class I YSOs, and relatively little mass is gained in the Class I, II, or III phases.

The idea that stars do not experience significant mass growth in the Class I, II, and III phases is consistent with what has previously been found regarding magnetospheric accretion. Figure 5 shows that the median mass accretion rate for our sample of T Tauri stars is $10^{-8}\ M_\odot\ \mathrm{yr}^{-1}$, and that the mass accretion rate falls below detection limits within about 5 Myr. In that time, of order 0.05 $M_\odot$ would be accumulated. This is consistent with the luminosity problem (S. J. Kenyon et al. 1990) for T Tauri stars, where their luminosity is too low for them to be accreting at a high enough rate to accumulate a solar mass of material in a few million years.

Episodic accretion has been suggested as a possible solution to the accretion problem. However, W. Fischer et al. (2023) show that a large EXor type outburst accumulates roughly a lunar mass of material, of order $10^{-8}\ M_\odot$. To build a solar-mass star, a YSO would need to be permanently in eruption, which is contradictory, for 20 times longer than the 5 Myr that mass accretion endures. FUors are another type of young eruptive variable, but with only 24 known as of 2018 (M. S. Connelley & B. Reipurth 2018) they are far too rare to account for significant mass growth for the majority of stars.[7] Episodic accretion events are unlikely to account for a significant fraction of a star's final mass.

If YSOs emerge as Class I objects near their final mass, then the vast majority of mass accretion must happen in the Class 0 phase, in less than 1 Myr. If true, then the accretion rate in the Class 0 phase must now be proportionally higher, and there is much less time to dissipate angular momentum. Regarding the accretion luminosity problem, rather than escaping as luminosity the potential energy of the infalling material can be sequestered as heat. An order-of-magnitude calculation shows that heating a solar mass of gas to $10^7$ K (roughly the core temperature of the Sun) requires the binding energy of the Sun. Regarding dissipating angular momentum, theory suggests that magnetic fields can remove up to one-half of the initial angular momentum (e.g., Y. Misugi et al. 2024). Fragmentation into binary stars may also be an effective way to solve the angular momentum problem by sequestering the angular momentum into the orbits of binary stars, and is consistent with the very high binary fraction in the Class 0 phase (X. Chen et al. 2013).

[7] The mass accretion rate during a FUor eruption is expected to be $10^{-4}\ M_\odot\ \mathrm{yr}^{-1}$ (L. Hartmann & S. J. Kenyon 1996). It takes $\sim 10^7$ yr for a star to get to the main sequence. At these mass accretion rates, a star would need to be in outburst for 0.1% of its YSO life to accumulate 1 $M_\odot$. The number of known YSOs is of order $10^{6.8}$ (c2D catalog), so there should be of order $10^{3.8}$ FUors in eruption at any given time if FUor eruptions are a significant mechanism for YSO mass growth. However, only 24 FUors are listed in M. S. Connelley & B. Reipurth (2018)

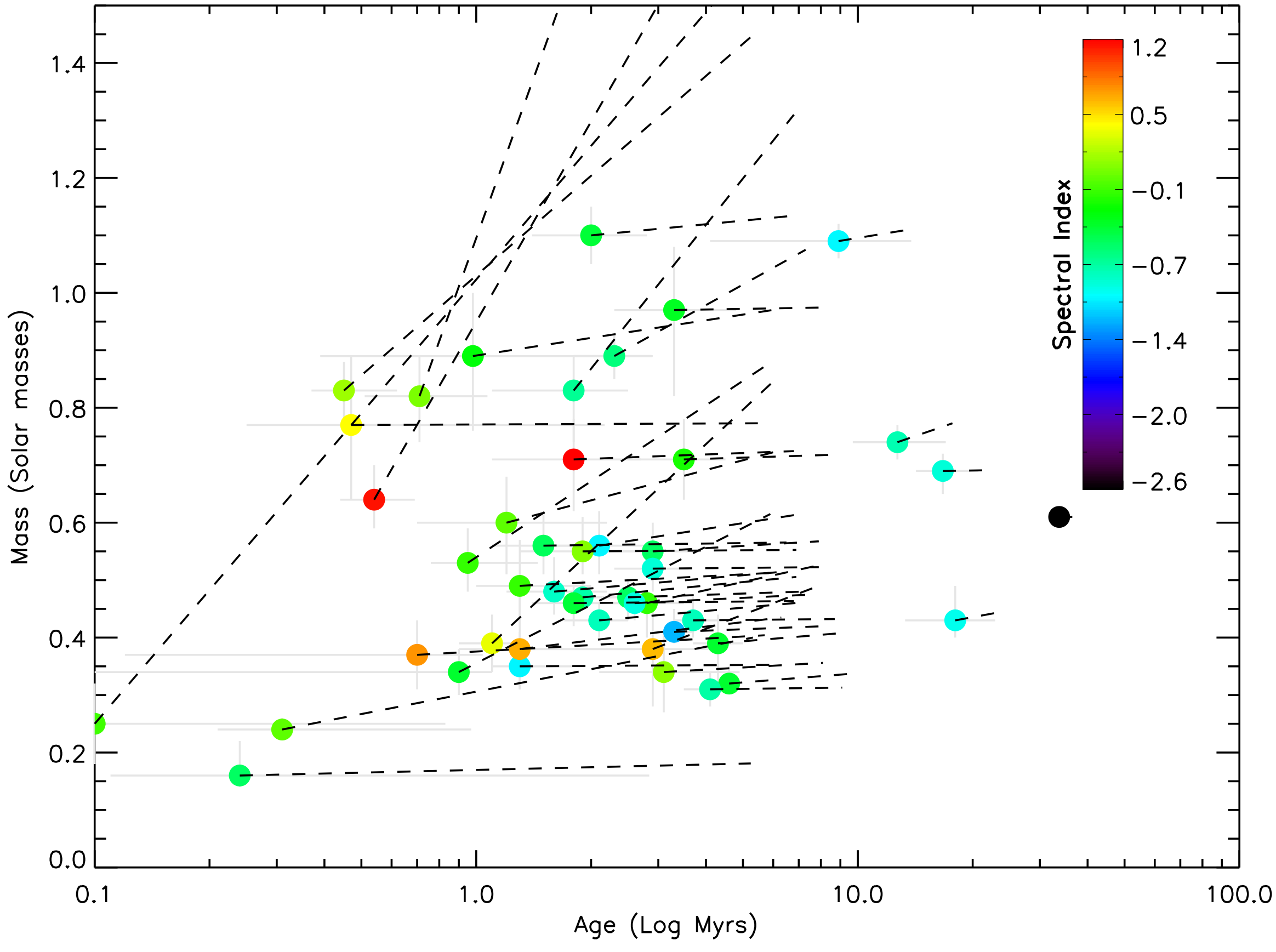

**Figure 7.** On this plot of mass vs. age, the dotted vectors show how much the mass of each YSO should increase given the current mass accretion rate in 5 Myr. Since mass accretion is variable, the accretion rate at the time of our measurement may not be representative of the accretion rate throughout that star's evolution. At the median mass accretion rate, ∼0.05 $M_{\odot}$ is expected to be gained in 5 Myr. Eruptive events either do not add enough mass or are too rare to significantly contribute to the growth of most stars, suggesting that young stars are nearly at their final mass by time they are visible in the near-IR.

### 5.4. The Evolution of a Young Stellar Object through Classes I, II, and III

Age is obviously one factor that determines whether an object is a Class I, II, or III YSO. However, considering the large range in ages at any given spectral index, and the range of spectral indices at a given age (Figure 2), age cannot be the only factor, and age may not be the most important factor. The departure of YSOs from their cloud cores certainly influences when a YSO evolves to a later class. Upon leaving the cloud of their birth, young stars would quickly become optically visible. Having been starved of a large gas reservoir, their IR excess and spectral index should decrease rapidly. However, if a YSO can remain in its natal cloud, then these objects might remain as an earlier class until their envelopes are exhausted, and be visible only in the IR due to the associated high extinction. Having access to a reservoir of material, they can maintain a high IR excess and a high spectral index. In this scenario, low-veiling Class I and Class II YSOs differ primarily on whether or not they have left their cloud. The ratio of Class I to Class II sources can then be interpreted as reflecting the probability of remaining in the cloud until an age of a few million years. The probability of a YSO having left its cloud increases with time, contributing to the higher fraction of Class II objects at later times.

When a YSO is ejected from its natal cloud, then it likely carries less of its envelope with it and thus would have lower extinction. If a Class I YSO could leave its cloud yet also keep a high IR excess for an extended period of time, then we would expect to observe a large number of objects with low extinction and a high spectral index. Figure 8 (top) shows that there are no such objects in our sample. We see that the extinction tends to increase with increasing spectral index. Extinction can affect the spectral index, but much of this trend remains after correcting the spectral index for extinction (Figure 8, bottom). The existence of this correlation between extinction-corrected spectral index and extinction is consistent with the scenario that the availability of material from the cloud is necessary to sustain a high spectral index, and that once that material is unavailable, the spectral index quickly declines.

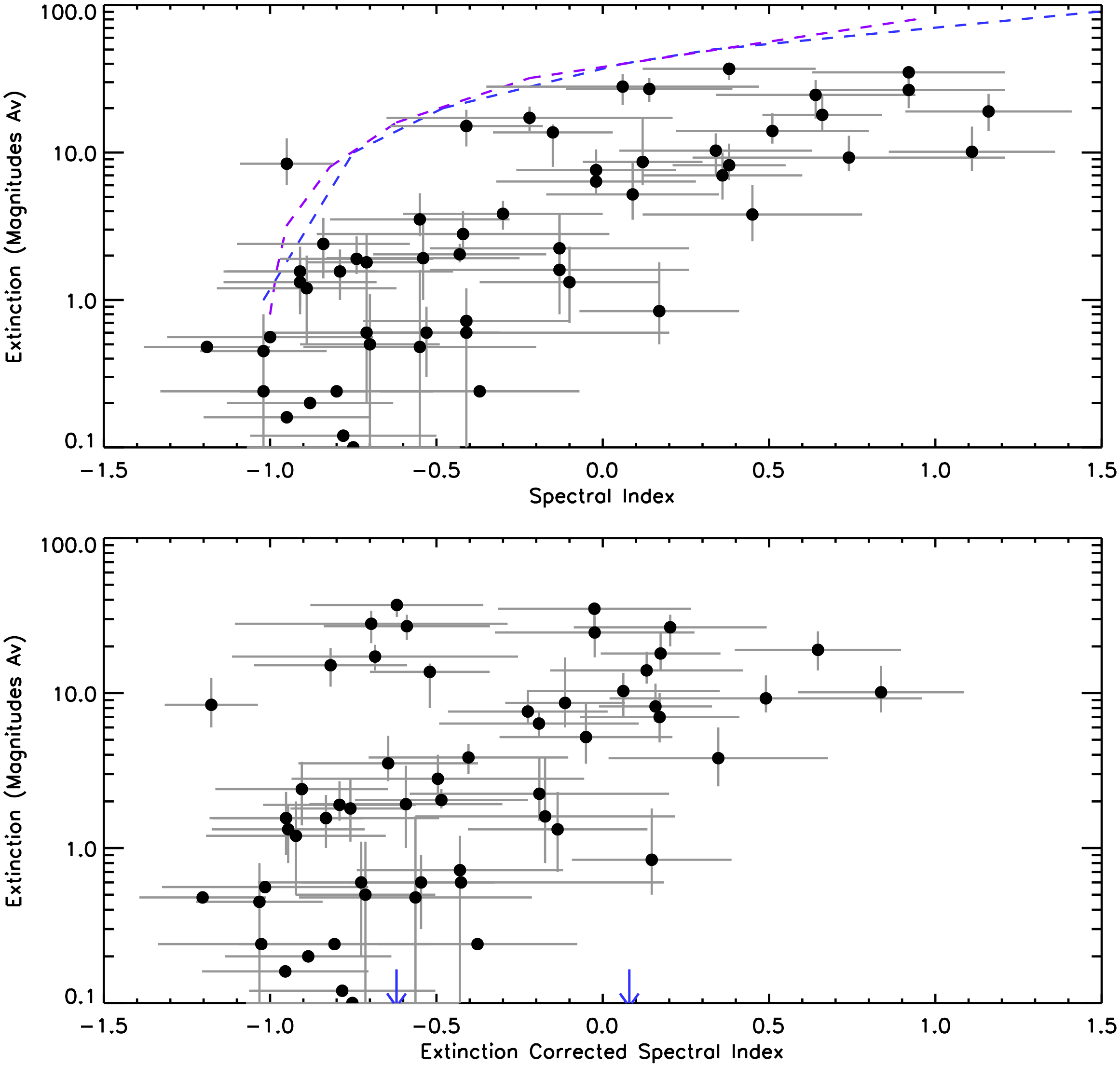

**Figure 8.** Top: the close correlation between magnitudes of visual extinction and spectral index, shown here for Class I and II YSOs, is in part due to extinction itself. The blue curve illustrates the path of a CTTS on this diagram as extinction is added, using the extinction law in A. N. Cox (2000), while the purple curve uses the extinction law from R. Indebetouw et al. (2005). Most objects have a higher spectral index than explained by extinction alone, showing that they have a true IR excess. We note that there are no objects with a high spectral index and low extinction, exactly what one would expect from a recently ejected YSO. This suggests that once ejected, the IR excess quickly decreases to Class II levels. More importantly, high extinction is necessary to maintain a high spectral index, i.e., an object needs to be embedded in a cloud to have a high IR excess. Bottom: after correcting the spectral indices for extinction, six deeply embedded Class I YSOs have corrected spectral indexes that are more consistent with a CTTS, suggesting that these may be good candidates for far-side T Tauri stars, or have edge-on disks.

### *5.5. Binarity*

While low-veiling Class I and Class II YSOs are physically very similar (C. Flores et al. 2024) they do have important differences, in particular with regards to binary fraction. M. S. Connelley et al. (2008), later confirmed by J. J. Tobin et al. (2022), showed that the binary fraction at wide separations steeply declines with spectral index. Hence, Class I YSOs (both high and low veiling) have a higher binary fraction than Class II YSOs, which in turn have a higher binary fraction than Class III YSOs. Later, X. Chen et al. (2013) have shown that Class 0 YSOs have an even higher binary fraction than Class I YSOs. These results were understood to show that the binary fraction decreases with time. How can we understand the change in binary frequency in light of the findings in C. Flores et al. (2024) and this paper that there is significant overlap in the ages of the low-veiling Class I, Class II, and Class III YSOs?

M. S. Connelley et al. (2008) found that the binary fraction of distant companions between 50 and 4500 au and the mean binary separation both declined with decreasing spectral index, and they suggested that this could be due to the gradual loss of the circumstellar envelope, which would otherwise help to retain loosely bound companions at wide separations. In Section 4.4, we suggested that a key differentiator between Class I and Class II YSOs is that the Class I YSOs have remained within their cloud whereas the Class II YSOs have left their cloud. These two ideas together suggest that Class I YSOs, having remained in their cloud, have held onto more circumstellar material that also helps to retain loosely bound binary companions. Class I YSOs may have a higher binary fraction not only because they are on average younger (and thus these systems are less likely to have dynamically decayed), but also because their greater envelope mass may have delayed the loss of binary companions that would otherwise have been lost if that circumstellar material were absent.

### 5.6. Far-side T Tauri Stars

It is likely that some Class I objects actually are T Tauri stars on the far side of the cloud and are seen through high extinction. M. M. Dunham et al. (2015) noted that a number of Class 0 + I YSOs have bolometric temperatures consistent with Class II or III objects, and conclude that these are likely to be more evolved YSOs seen through high extinction. Although T Tauri stars are usually optically visible, we would expect there to be as many T Tauri stars on the far sides of molecular clouds as on the near sides. The similarity of the apparent ages (via the log($g$) distributions) of low-veiling Class I and Class II YSOs opens the possibility that some Class I YSOs indeed could be T Tauri stars on the far side of the cloud.

The median spectral index of our sample of T Tauri stars is roughly −1. To make a T Tauri star look like a Class I YSO, ∼40 mag of visual extinction would be required to change the spectral index (calculated from 2 to 22 $\mu$m) to be greater than zero. This is more extinction than the median extinction of Class I YSOs ($A_V \sim$ 15–20), but within the maximum extinction for our sample of Class I YSOs.

Figure 8 (bottom) shows that six out of 32 YSOs with $\alpha > -0.5$ have a spectral index more consistent with a Class II YSO ($\alpha < -0.5$) after correcting for extinction. This suggests that, for these objects, extinction is the main reason they have a high spectral index, and that the star has little intrinsic IR excess. Such objects could be far-side T Tauri stars and, if so, it suggests that 19% ± 8% of flat-spectrum and Class I YSOs could be far-side T Tauri stars.

Two of these stars have Gaia parallaxes, and both happen to be in the $\rho$ Oph star-forming region. Their distances are 116 ± 13 pc (GSS 39) and 153 ± 3 pc (IRAS 16288-2450 W2). GSS 39 appears to be associated with L1688, which is 134 ± 5 pc away C. Zucker et al. (2020), and hence would be in front of it. IRAS 16288-2450 W2 appears to be associated with B44, which is 154 ± 7 pc away C. Zucker et al. (2020), and hence would be coincident with it.

### 5.7. Potential Impact of High-veiling Class I Young Stellar Objects

In this analysis we have not been able to include Class I YSOs with high veiling because high veiling precludes observing photospheric absorption lines. Roughly one-half of Class I YSOs have high veiling, and high-veiling Class I YSOs have a higher average spectral index than low-veiling Class I YSOs. We do not believe that the addition of the high-veiling Class I YSOs would substantially affect the main conclusions of this paper. For example, Figure 8 shows that there are no high-index objects with low extinction, supporting the hypothesis that the spectral index of ejected YSOs quickly decline (Section 4.4). High-veiling Class I YSOs also have high extinction, and would not affect this conclusion. The overlap in age, mass, and other properties between low-veiling Class I and Class II YSOs would remain even after the addition of high-veiling Class I YSOs.

We do not expect the physical properties of high-veiling Class I YSOs to be significantly different than low-veiling Class I YSOs as veiling itself is variable. M. S. Connelley & T. P. Greene (2014) found that veiling was observed to be variable in each object for which they could measure veiling, and one object (IRAS 03301+3111) was observed to transition from low to high veiling. During the course of this study, we attempted to reobserve many of the targets in G. W. Doppmann et al. (2005). Whereas all of the targets in common with G. W. Doppmann et al. (2005) had photospheric lines when observed in 2003, roughly one-quarter did not by the time we observed them in 2016 and later, presumably because the veiling had increased between observations. The underlying object is physically the same despite the changes in the veiling in these examples. While high veiling may be more common at younger ages, we expect high-veiling Class I YSOs to be physically similar to their low-veiling counterparts.

## 6. Conclusions

We have analyzed the stellar parameters for a large sample of low-veiling Class I, Class II, and Class III YSOs derived from high-resolution near-IR spectroscopy, using stellar surface gravity as a tracer of age that is independent of the circumstellar environment. We compared these gravities to masses, extinctions, mass accretion rates, rotation, and veiling derived from near-IR spectroscopy and archival photometry in order to explore how these parameters evolve with time. Since we used the same instruments, observing techniques, data reduction tools, stellar atmospheric modeling codes, and analysis tools for all of our objects, we can directly compare the properties of our large sample of these three classes of YSOs. For Class I objects we were only able to determine the physical parameters of objects with veiling low enough to allow us to see the photospheric lines, and this limitation is important to keep in mind while interpreting these results. Our key findings are summarized in the following.

(1) *Spectral index*. While the spectral index on average decreases with increasing age, there is a large scatter of approximately ±1. Class I, II, and III YSOs can all be the same age. In addition to the IR excess, extinction also affects the spectral index, further complicating using the spectral index as an age proxy. Correcting for extinction causes some Class I YSOs to fall among the Class II YSOs.

(2) *Extinction.* YSOs can have a wide range of extinctions at a given age. For example, at log($g$) ≈ 3.4–3.6, the extinctions range from optically visible ($A_V \approx 0$) to deeply embedded ($A_V \approx 40$) over this range of gravities. Thus, extinction is a poor indicator of age. We also do not find a correlation between age and extinction for Class I and II YSOs, whereas YSOs older than ∼5 Myr all have low extinction. Some objects in our sample are ∼1 Myr old but are already optically visible, while others are ∼3 Myr old yet are still deeply embedded.

(3) *Mass accretion rate*. Magnetospheric accretion, traced by the Br$\gamma$ line flux, decreases with time, but the correlation is weak with a scatter of ∼1 dex. There appears to be an upper threshold of mass accretion rate that decreases with time. Accretion effectively ends (decreases below a threshold of $10^{-10}\ M_\odot\ \mathrm{yr}^{-1}$) for most stars by the time log($g$) exceeds 4.0, at approximately 5 Myr. Magnetospheric accretion is not expected to significantly add to the final mass of the star. As eruptions are inadequate to accumulate a significant fraction of a star's mass, stars are likely to be near their final mass by the time they are detectable in the IR as Class I YSOs.

(4) *Class II YSOs are physically very similar to low-veiling Class I YSOs*. They have similar mass accretion rates and overlapping gravity ($\propto$age) and temperature ($\propto$mass) distributions. The 0.8–2.4 $\mu$m medium-resolution spectra of Class II and low-veiling Class I YSOs are generally very similar, except for changes in slope due to extinction.

## Acknowledgments

We are grateful for the professional assistance from Dave Griep, Brian Cabreira, Greg Engh, Tony Matulonis, Greg Osterman, and Bernie Walp. This research has made use of the SIMBAD database, operated at CDS, Strasbourg, France, and NASA's Astrophysics Data System. This publication makes use of data products from the Two Micron All Sky Survey, which is a joint project of the University of Massachusetts and the Infrared Processing and Analysis Center/California Institute of Technology, funded by the National Aeronautics and Space Administration and the National Science Foundation. This research has made use of NASA's Astrophysics Data System, operated by the Smithsonian Astrophysical Observatory under NASA Cooperative Agreement 80NSSC21M0056. This publication makes use of data products from the Wide-field Infrared Survey Explorer, which is a joint project of the University of California, Los Angeles, and the Jet Propulsion Laboratory/California Institute of Technology, funded by the National Aeronautics and Space Administration. This work has made use of data from the European Space Agency (ESA) mission Gaia,[8] processed by the Gaia Data Processing and Analysis Consortium (DPAC).[9] Funding for the DPAC has been provided by national institutions, in particular the institutions participating in the Gaia Multilateral Agreement. Christian Flores acknowledges support from ANID–Millennium Science Initiative Program—Center Code NCN2024_001.

*Facility:* IRTF.

## ORCID iDs

Michael Connelley https://orcid.org/0000-0002-8293-1428
Christian Flores https://orcid.org/0000-0002-8591-472X

[8] https://www.cosmos.esa.int/gaia
[9] https://www.cosmos.esa.int/web/gaia/dpac/consortium